\documentclass[a4paper,fleqn]{cas-sc}

\usepackage[numbers]{natbib}
\usepackage{mathtools}
\usepackage[scientific-notation=true]{siunitx}

\usepackage{framed} 
\usepackage{multicol} 

\makeatletter
\def\@seccntformat#1{\@ifundefined{#1@cntformat}%
   {\csname the#1\endcsname\quad}
   {\csname #1@cntformat\endcsname}
}
\makeatother
 
\usepackage{nomencl} 
\usepackage[boxruled]{algorithm2e} 
\makenomenclature
\renewcommand*\nompreamble{\begin{multicols}{2}}
\renewcommand*\nompostamble{\end{multicols}}

\renewcommand\nomgroup[1]{%
  \item[\bfseries
  \ifstrequal{#1}{A}{Abbreviations}{%
  \ifstrequal{#1}{B}{Basic definitions}{%
  \ifstrequal{#1}{V}{Variables}{%
  \ifstrequal{#1}{P}{Parameter}{%
  \ifstrequal{#1}{S}{Sets}{%
  \ifstrequal{#1}{F}{Functions and mappings}}}}}}%
]}

\begin{document}
\let\WriteBookmarks\relax
\def\floatpagepagefraction{1}
\def\textpagefraction{.001}
\shorttitle{Graph-based formulation for macro-energy systems}
\shortauthors{Göke}

\title [mode = title]{A graph-based formulation for modeling macro-energy systems}

\author[1]{Leonard Göke}[type=editor,auid=000,bioid=1]
\cormark[1]
 \ead{lgo@wip.tu-berlin.de}

\address[1]{Berlin University of Technology, Workgroup for Infrastructure Policy (WIP), Berlin, Germany.}

\cortext[cor1]{Corresponding author.}

\begin{abstract}
Averting the impending harms of climate change requires to replace fossil fuels with renewables as a primary source of energy. Non-electric renewable potential being limited, this implies extending the use of electricity generated from wind and solar beyond the power sector, either by direct electrification or synthetic fuels. Modeling the transformation towards such an energy system is challenging, because it imposes to consider fluctuations of wind and solar and the manifold ways the demand side could adjust to these fluctuations.

This paper introduces a graph-based method to formulate energy system models to address these challenges. By organizing sets in rooted trees, two features to facilitate modeling high shares of renewables and sector integration are enabled. First, the method allows the level of temporal and spatial detail to be varied by energy carrier. This enables modeling with a high level of detail and a large scope, while keeping models computationally tractable. Second, the degree to which energy carriers are substitutable when converted, stored, transported, or consumed can be modeled to achieve a detailed but flexible representation of sector integration. An application of the formulation demonstrates that the variation of temporal detail achieves an average reduction in computation time of 70\%.

\end{abstract}

\begin{highlights}
\item A graph-based formulation for modelling energy-systems with high levels of renewables and sector integration is introduced.
\item The method allows the level of temporal and spatial detail to be varied by energy carrier. This enables modeling with a high level of detail, while keeping models computationally tractable.
\item The degree to which energy carriers are substitutable when converted, stored, transported, or consumed can be modeled.
\end{highlights}


\maketitle

\begin{table*}[pos=!t]
   \begin{framed}
     \printnomenclature[1cm]
   \end{framed}
\end{table*}

\nomenclature[A, 01]{CCGT}{combined-cycle gas turbine}
\nomenclature[A, 02]{CHP}{cogeneration of heat and power}

\nomenclature[B, 01]{$rt_{G}$}{root of tree $G$}
\nomenclature[B, 02]{$\alpha_{v},\, \alpha_{v}^{+}$}{ancestors of vertex $v$, $+$ includes $v$}
\nomenclature[B, 03]{$\alpha_{v}^{z}$}{ancestors of vertex $v$ at depth $z$}
\nomenclature[B, 04]{$\delta_{v},\, \delta_{v}^{+}$}{descendants of vertex $v$, $+$ includes $v$}
\nomenclature[B, 05]{$\delta_{v}^{z}$}{descendants of vertex $v$ at depth $z$}
\nomenclature[B, 06]{$\lambda_{v}$}{leaves descendant to $v$}

\nomenclature[V, 01]{\makebox[45pt][l]{$Te_{\hat{t},\hat{r},\hat{c}}^{cv/st}$}}{net-output of conversion/storage}
\nomenclature[V, 02]{\makebox[45pt][l]{$Exc_{\hat{t},\hat{r},\hat{c}}^{net}$}}{net-exchange}
\nomenclature[V, 03]{\makebox[45pt][l]{$Trd_{\hat{t},\hat{r},\hat{c}}^{net}$}}{net-trade}
\nomenclature[V, 04]{\makebox[45pt][l]{$Cv_{t,\tilde{t},r,c,e,m}^{in/out}$}}{aggregated conversion input/output}
\nomenclature[V, 05]{\makebox[45pt][l]{$St_{t,\tilde{t},r,c,e,m}^{in/out}$}}{aggregated storage input/output}
\nomenclature[V, 06]{\makebox[45pt][l]{$Gen_{t,\tilde{t},r,c,e,m}$}}{generated energy}
\nomenclature[V, 07]{\makebox[45pt][l]{$Use_{t,\tilde{t},r,c,e,m}$}}{used energy}
\nomenclature[V, 08]{\makebox[45pt][l]{$StO_{t,\tilde{t},r,c,e,m}^{ext/int}$}}{externally/internally discharged energy}
\nomenclature[V, 09]{\makebox[45pt][l]{$StI_{t,\tilde{t},r,c,e,m}^{ext/int}$}}{externally/internally charged energy}
\nomenclature[V, 10]{\makebox[45pt][l]{$StLvl_{t,\tilde{t},r,c,e,m}$}}{storage level}
\nomenclature[V, 11]{\makebox[45pt][l]{$Exc_{t,r,r',c}$}}{energy exchange from region $r$ to region $r'$}
\nomenclature[V, 12]{\makebox[45pt][l]{$Trd_{t,r,c,i}^{buy/sell}$}}{bought/sold energy}
\nomenclature[V, 13]{\makebox[45pt][l]{$Cap_{t,\tilde{t},r,e}^{opr/ist,cv}$}}{operated/installed conversion capacity}
\nomenclature[V, 14]{\makebox[45pt][l]{$Cap_{t,\tilde{t},r,e,c}^{opr/ist,stI}$}}{operated/installed storage input capacity}
\nomenclature[V, 15]{\makebox[45pt][l]{$Cap_{t,\tilde{t},r,e,c}^{opr/ist,stO}$}}{operated/installed output storage capacity}
\nomenclature[V, 16]{\makebox[45pt][l]{$Cap_{t,\tilde{t},r,e,c}^{opr/ist,stS}$}}{operated/installed storage size}
\nomenclature[V, 17]{\makebox[45pt][l]{$Cap_{t,r,r',c}^{opr/ist,exc}$}}{operated/installed exchange capacity}
\nomenclature[V, 18]{\makebox[45pt][l]{$Exp_{t,r,e}^{cv}$}}{expansion of conversion capacity}
\nomenclature[V, 19]{\makebox[45pt][l]{$Exp_{t,r,c,e}^{stI}$}}{expansion of storage input capacity}
\nomenclature[V, 20]{\makebox[45pt][l]{$Exp_{t,r,c,e}^{stO}$}}{expansion of storage output capacity}
\nomenclature[V, 21]{\makebox[45pt][l]{$Exp_{t,r,c,e}^{stS}$}}{expansion of storage size}
\nomenclature[V, 22]{\makebox[45pt][l]{$Exp_{t,r,r',c}^{exc}$}}{expansion of exchange capacity}

\nomenclature[P, 01]{\makebox[45pt][l]{$ava_{t,\tilde{t},r,e,m}^{cv}$}}{availability of conversion capacity}
\nomenclature[P, 02]{\makebox[45pt][l]{$ava_{t,\tilde{t},r,c,e,m}^{stI/stO/stL}$}}{availability of storage capacity}
\nomenclature[P, 03]{\makebox[45pt][l]{$eff_{t,\tilde{t},r,e,m}^{cv}$}}{efficiency of conversion process}
\nomenclature[P, 04]{\makebox[45pt][l]{$eff_{t,\tilde{t},r,c,e,m}^{stI/stO}$}}{efficiency of charging/discharging}
\nomenclature[P, 05]{\makebox[45pt][l]{$eff_{t,r,r',c}^{exc}$}}{efficiency of energy exchange}
\nomenclature[P, 06]{\makebox[45pt][l]{$ratio_{t,\tilde{t},r,c,e,m}^{out, eq}$}}{fixed share of carrier $c$ on total output}
\nomenclature[P, 07]{\makebox[45pt][l]{$in_{t,\tilde{t},r,c,e,m}$}}{inflows into storage system}
\nomenclature[P, 08]{\makebox[45pt][l]{$dis_{t,\tilde{t},r,c,e,m}$}}{self-discharge rate of storage}
\nomenclature[P, 09]{\makebox[45pt][l]{$dem_{\hat{t},\hat{r},\hat{c}}$}}{energy demanded}
\nomenclature[P, 10]{\makebox[45pt][l]{$cap_{t,r,c,i}^{buy/sell}$}}{capacity for buying/selling}

\nomenclature[S, 01]{$\Omega$}{all possible indices for dispatch variables of technologies}
\nomenclature[S, 02]{$\Gamma^{cv/st}$}{technologies converting/storing carriers}
\nomenclature[S, 03]{$\Psi_{e}^{in/out}$}{pairs defining capacity constraints on conversion input/output}
\nomenclature[S, 04]{$\gamma_{e}^{use/gen}$}{carriers used/generated by technology $e$}
\nomenclature[S, 05]{$\gamma_{e}^{stEx}$}{carriers stored explicitly and externally}
\nomenclature[S, 06]{$\gamma_{e}^{stCap}$}{carriers assigned to storage capacity}
\nomenclature[S, 07]{$\gamma_{e}^{st}$}{all carriers stored explicitly}
\nomenclature[S, 08]{$\gamma_{e}^{in/out}$}{external input/output carriers}
\nomenclature[S, 09]{$\mu_{e}$}{modes assigned to technology $e$}
\nomenclature[S, 10]{$\tau_{c}$}{dispatch time-steps}
\nomenclature[S, 11]{$\rho_{c}$}{dispatch regions}
\nomenclature[S, 12]{$\varphi_{c}$}{pairs of dispatch time-steps and regions}
\nomenclature[S, 13]{$\sigma_{\hat{c},r,t}$}{pairs of dispatch time-steps and regions aggregated to determine dispatch of $\hat{c}$ at time-step $t$ in region $r$}
\nomenclature[S, 14]{$\epsilon_{e}$}{pair of dispatch time-steps and regions the conversion balance is created for}
\nomenclature[S, 15]{$\theta_{e,\tilde{t}}^{dis}$}{time-steps of construction considered dispatched separately}
\nomenclature[S, 16]{$\theta_{e,t,\tilde{t}}^{exp}$}{time-steps of construction aggregated to obtain capacity}
\nomenclature[S, 17]{$\beta_{c,r}$}{regions with that region $r$ can exchange carrier $c$}
\nomenclature[S, 18]{$\zeta_{t,r,c,i}^{buy/sell}$}{steps in supply/demand curve for trade}
\nomenclature[S, 19]{$\omega_{t,\tilde{t},r,e}^{cv/st}$}{set of modes, each set requires an individual conversion/storage balance }
\nomenclature[S, 20]{$\eta_{e}^{tp/sp}$}{time-steps/regions of capacity expansion}

\nomenclature[F, 01]{$d(v)$}{depth of vertex $v$}
\nomenclature[F, 02]{$dep_{c}$}{depth assigned to carrier $c$}
\nomenclature[F, 03]{$dep_{sup}$}{depth of superordinate dispatch time-steps}
\nomenclature[F, 04]{$s(t)$}{scaling factor for capacities at time-step $t$}
\nomenclature[F, 05]{$g(e)$}{type assigned to technology $e$}

\section{Introduction} \label{01}

Averting the impending harms of climate change requires to cut carbon emissions from current record highs to zero by at least 2050. Fossil fuels account for the three quarters of all emissions and consequently need to be replaced by renewable energies \citep{IPCC2014}. Especially wind and solar have to take a predominant role, since their unexploited potential greatly exceeds hydro or biomass. 

This transformation has profound implications for the entire energy system: On the supply side, the fluctuating nature of wind and solar requires additional flexibility to be reliable. On the demand side, the source of primary energy must shift towards electricity from wind and solar, either by direct electrification or synthetic fuels. To provide flexibility and shift primary energy to renewable electricity, the different sectors of the energy system have to be closely integrated. Charging electric vehicles for instance depends on supply from the power sector, but can also contribute to balancing fluctuating supply with demand \citep{Doucette2011}. Similarly, many industrial processes require renewable electricity for decarbonization, but are capable of adding flexibility too \citep{Burre2020}.

Overall, these profound changes of the energy system result in new demands on models analyzing and planning energy systems. To address these demands, \citet{Levi2019} propose the discipline of "macro-energy systems" that is characterized by a large scope, covering several years, different sectors, and a large region and, as a consequence, a high level of complexity, that necessitates great abstraction. Following up on this idea, \citet{DeCarolis2020} argue that the challenges in modelling marco-energy system can best be overcome by collaborative development of open-source tools. The call for openness is also prominent in other publications and is a consequence of the impact models can have on energy and climate policy, since they allow assessing alternative designs of the system in terms of costs and emissions \citep{Pfenninger2017b,Weibezahn2019}.

This paper introduces a novel graph-based formulation for modelling macro-energy systems. This novel formulation specifically addresses the challenges that the transformation towards a system with high levels of renewables and sector integration imposes. The following literature review provides a detailed overview of these challenges and how existing modelling frameworks meet them. Afterwards, section \ref{1} presents the graph-based formulation and its distinctive features by listing the sets and equations constituting the model's underlying optimization problem. In section \ref{4} the formulation is applied to create an example model and demonstrate the benefits of the introduced formulation. For this purpose, the open-source modelling framework AnyMOD.jl that implements the graph-based formulation is used. Finally, section \ref{5} concludes.

\section{Literature review} \label{0}

Subsection \ref{11} summarizes the technical challenges in modelling future energy systems that previous research identified. The following subsection discusses how existing modelling frameworks address these challenges.

\subsection{Challenges in macro-energy system modeling} \label{11}

A key requirement when modeling energy systems with large shares of renewables is high temporal granularity \citep{Pfenninger2014}. Former research shows that the number of representative time-steps an entire year can be reduced to strongly depends on the share of weather-dependant generation. At low resolutions utilization of wind and solar is overestimated, since fluctuations of supply cannot be captured adequately \citep{Poncelet2016, Nahmacher2014, Haydt2011}. Reinforced sector integration may cause a similar effect on electricity demand, if heat supply is increasingly electrified by electrical heat pumps \citep{Bloess2019}. Since all these temporal fluctuations are weather related and thus subject to uncertainty, high temporal granularity is ideally combined with a stochastic approach \citep{Ringkjob2018}.

At the same time, spatial aspects gain in relevance too, when modeling high levels of renewables, since their \textit{“economic potential and generation costs depend greatly on their location”}  \citep{Pfenninger2014}. In addition, in a renewable system the capacity of individual generation units is about a magnitude smaller than in a system characterized by thermal plants. This creates the opportunity to match demand with local supply as an alternative to transporting energy carriers over long distances \citep{Bauknecht2020}. However, modeling such solutions does not only require a consistent representation of relevant technologies, for instance solar home systems with batteries, but a high spatial granularity as well.

The need for temporal and spatial granularity when modeling high levels of intermittent renewables and sector integration is directly related to the concept of flexibility. Flexibility can be defined as an energy system's capability to cope with variability and uncertainty in demand and generation \citep{Heggarty2019}. The arising need for flexibility and how it can be satisfied is widely recognized as a key question for future energy systems \citep{Kondziella2016, Lund2015}. To fully account for these flexibility needs within models means to fully capture weather-driven fluctuations and consequently requires high temporal and spatial granularity.

On the other hand, including all options to provide flexibility into models calls for a detailed representation of sector integration. Many potential sources of flexibility involve complex interaction of technologies and energy carriers to build synergies between sectors \citep{Orths2019}. To give but one example, synthetic gas can be generated from electricity via electrolysis and methanation, when supply from wind or solar exceeds demand, stored and then used to provide heat or electricity at times of low intermittent supply. Models that omit these cross-sectoral sources of flexibility might fail to identify cost-efficient solutions and excessively invest into other storage and transport capacities instead \citep{Brown2018}.

Besides these challenges concerning granularity and detail, the way models are practically applied creates additional challenges that concern their temporal and spatial scope. Ideally, models can analyze how today's energy system can be transformed to comply with the climate objectives set for a certain year \citep{Oberele2019}. Therefore, their temporal scope should include multiple subsequent periods that are simultaneously optimized, also referred to as perfect foresight. If models are limited to single years, computing pathways has to rely on consecutively solving each year separately. This approach has been termed myopic foresight and found to cause suboptimal results due to stranded investments \citep{Loeffler2019, Gerbaulet2019}. A large spatial scope is valuable, because energy systems of different regions are increasingly interlinked, be it through a common energy policy or interconnected markets and networks, as for example in the European gas and electricity sector. The latter is again relevant from a flexibility perspective as well: Especially exchange of electric between regions, can even out local fluctuations of wind and solar generation \citep{Thellufsen2017}.

\subsection{How challenges are addressed} \label{12}

Former research already proposed several formulations for modelling energy systems. Typically, these formulations are embedded into a corresponding software tool, also referred to as modelling framework, that is used to generate specific models \citep{Groissboeck2019, Wiese2018}. In the following, two modelling frameworks, OSeMOSYS and Calliope, are evaluated with regard to the challenges outlined in section \ref{11} \citep{Howells2011, Pfenninger2018, Calliope2020}. The choice fell on these, because both are representative for a larger group of frameworks and models. OSeMOSYS is closely related to many long-established tools for energy system planning like PRIMES, MESSAGE or MARKAL. The Calliope framework draws parallels to more novel tools like Balmorel, PyPSA and DIETER that are more focused on the power sector and high accuracy regarding intermittent renewables \citep{Lopion2018, Groissboeck2019}.

These different contexts are reflected in the way OSeMOSYS and Calliope treat time, which again affects temporal granularity. OSeMOSYS pursues an approach that aggregates an entire year into a few representative periods, also referred to as time-slices (e.g. a summer evening). Modeling these periods instead of the full year greatly decreases computational effort, but also limits temporal granularity and thus the capability to capture fluctuations of intermittent renewables. To avoid this, Calliope does not rely on representative periods, but rather uses unaltered continuous time series.\footnote{Using representative periods is possible as well but is not the default option.} This comes at the cost of a steep increase in size and solve time, if not only the electricity sector, but the entire energy system is modeled \citep{Lopion2018}. Neither of the two frameworks can account for uncertainty of supply and demand.

The use of representative periods within OSeMOSYS also implies a loss of chronology, and thus restricts the modeling of storage, especially seasonal storage. If a time span, for example the entire summer, is reduced to one representative period, say a week, storage patterns determined for that period apply to the entire time span. Accordingly, storage levels would show the same pattern for each summer week and could not continuously increase over the course of the summer \citep{Welsch2012}. Since energy systems with high shares of variable renewables can be expected to heavily rely on storage, without major adjustments the approach is ill-suited to describe these systems \citep{Kotzur2018}.

The temporal scope of OSeMOSYS may include multiple subsequent periods of capacity expansion to compute long-term pathways for transforming the energy system. However, properties of technologies cannot depend on their respective period of construction. As a result, technological advances, like increasing efficiency of power-to-gas technologies for instance, cannot be accounted for adequately. Calliope is limited to a single period of capacity expansion.

Besides these differences regarding temporal detail, both frameworks can achieve high spatial granularity and a large regional scope, since the number of regions can be chosen freely. In addition, Calliope also provides an option for discrete expansion and dispatch of technologies, which renders it appropriate for applications as detailed as the building level.

To extend the representation of technologies, OSeMOSYS supports different modes of operation, like either operating a CHP plant at a higher fuel utilization rate, but a smaller CHP coefficient or the other way round. Calliope provides a functionality to include technologies that can store a carrier for later use within a conversion, for example concentrated solar power plants that store heat for later conversion into electricity.

\section{Model formulation} \label{1}

The graph-based formulation introduced in this paper relies on continuous time series instead of representative periods to achieve the temporal detail necessary for high shares of variable renewables. To model the long-term transformation of the energy system, it supports multiple periods of capacity expansion accounting for technological advance and endogenous decommissioning of capacities. In addition, it extends Calliope’s functionality for technologies that first store and later use a carrier by also allowing for technologies that first generate and then store a carrier. This enables modeling decentralized storage systems, like a home battery paired with a photovoltaic panel, within large-scale system models. Furthermore, different operational modes for technologies are supported.

Beside these gradual improvements, the proposed formulation introduces two novel features to facilitate modelling high levels of renewables and sector integration:

\begin{enumerate} 
\itemsep=0pt
\item The level of temporal and spatial granularity can be varied by energy carrier. For instance, electricity can be modeled with hourly resolution, while supply and demand of gas is balanced daily. This achieves the temporal granularity required to capture fluctuations in renewable electricity generation but avoids applying it to all other carriers as well. Within spatially aggregated models, for many carriers, like gas for instance, a less detailed resolution will better reflect physical properties and avoid inflating the model. In \citet{Renaldi2017}, a similar method from process system engineering is used to optimize a solar district heating system. However, the method is only applied to the temporal granularity only and to technologies instead of carriers.
\item Substitution of energy carriers can be modeled in dependence of the respective context: conversion, storage, transport or demand. For example, heat from residential heat pumps and district heating plants might both satisfy heat demand, but only district heat can be stored within large-scale storage systems.
\end{enumerate}  

Since the proposed formulation is specifically aimed at macro-energy systems, it does not support discrete expansion and dispatch of technologies and therefore is not suited to be applied at the urban or building level. 

\subsection{Sets and mappings} \label{2}

This section discusses the sets defined within the modelling framework, in particular time-steps, regions, energy carriers, technologies, and modes, and how they are mapped to each other. To facilitate comprehension, the whole introduction of the framework revolves around an example model. Since the primary interest of that model is not its specific results, but its general method, the choice of energy carriers and technologies considered is not exhaustive. For the same reason, some modeling assumptions that could be argued to require an in-depth technical discussion, are only treated briefly.

Since the framework organizes all sets within rooted trees, first some concepts of graph theory and basic notations used throughout the paper have to be introduced. Any graph $G$ is defined by its vertices $V$ and edges $E$. A tree can be defined as a graph, where any two vertices are linked by a unique path along its vertices and edges. Distinguishing one vertex as the graph's root $rt_{G}$ creates a rooted tree. The length of a path from a vertex $v$ to the root is termed depth and provided by function $ d{:V} \to \mathbb{N}_{0}$. Consequently, the depth of the root is always zero, which means that $ d(rt_{G}) = 0$. All vertices on the path between a vertex $v$ and the root are its ancestors and defined as set $\alpha_{v}$. The descendants of a vertex $v$, henceforth given as $\delta_{v}$, can be understood recursively: If a vertex $u$ is an ancestor to $v$, $v$ is a descendant to $u$. To indicate the vertex $v$ itself should be included in a set of ancestors or descendants, we write $\alpha_{v}^{+}$ or $\delta_{v}^{+}$, respectively. The set of all ancestors or descendants of vertex $v$ with depth $z$ is denoted as $\alpha_{v}^{z}$ and $\delta_{v}^{z}$. A subgraph of a tree that only contains the vertex $v$ and all its descendants, is referred to as the subtree $G_{v}$. Lastly, all vertices without any descendants are called leaves. For all leaves, which are descendants of vertex $v$, we write $\lambda_{v}$. \citep{Diestel2000, Bondy2008}

\subsubsection{Regions}

\begin{figure}
	\centering
		\includegraphics[scale=.5]{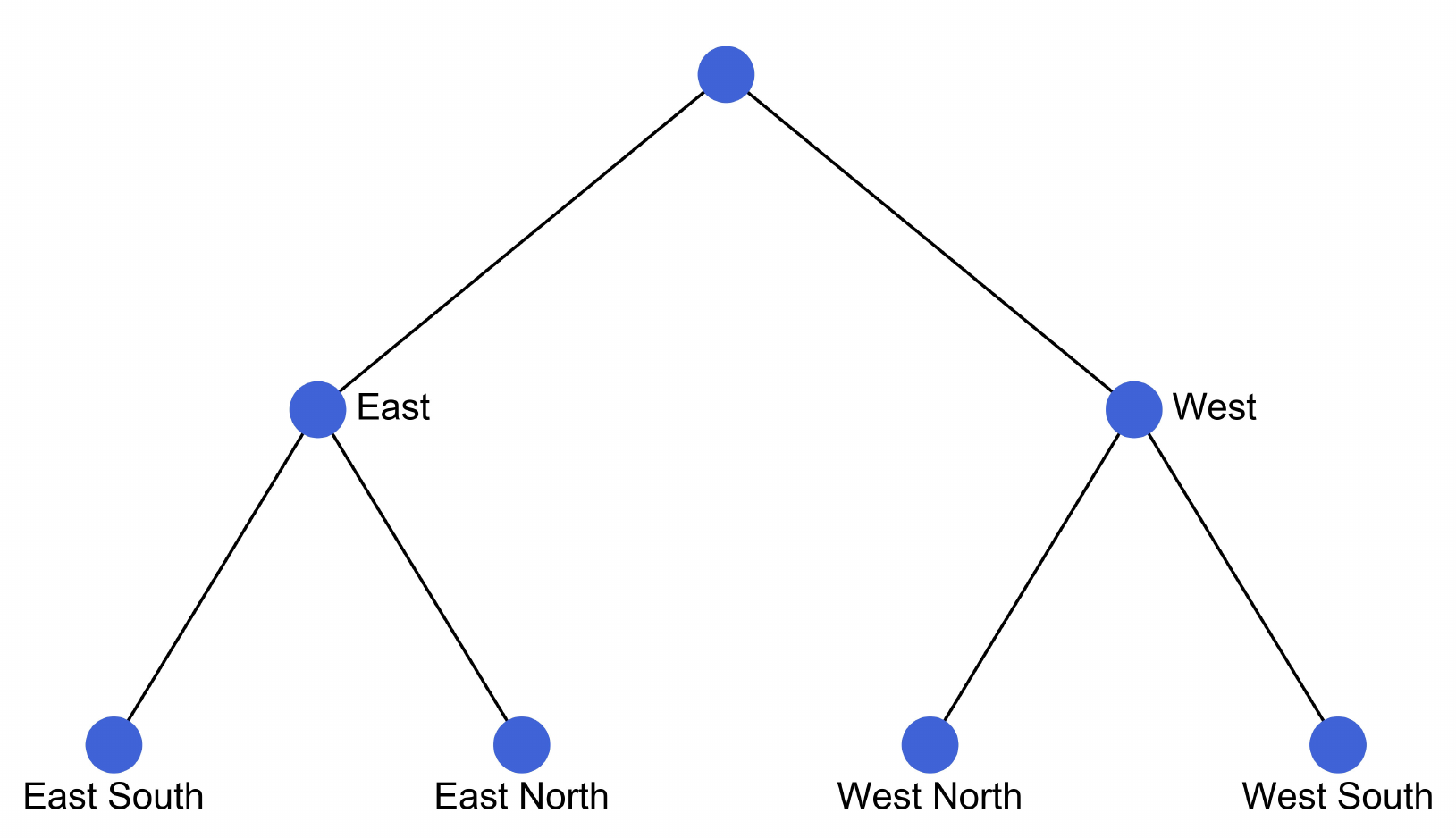}
	\caption{Rooted tree of regions in the example model}
	\label{fig:1}
\end{figure}

Fig. \ref{fig:1} shows the rooted tree $R$ organizing all regions considered within the example problem. $r$ be an arbitrary vertex of the tree representing a region. Exemplifying the definitions and notations introduced above, the descendants of vertex \textit{'East'} are the vertices \textit{'East South'} and \textit{'East North'} or $\delta_{East} = \{\textit{'East South'},\textit{'East North'}\}$. Since both \textit{'East South'} and \textit{'East North'} do not have any descendants, they are leaves and $\lambda_{East} = \delta_{East}$ applies. Also, the ancestor of vertex \textit{'West North'} at depth 1 is the vertex \textit{'West'}, which means $\alpha_{West North}^{1} = \{\textit{'West'}\}$. The subtree at vertex \textit{'West'} would include the vertices \textit{'West'}, \textit{'West North'} and \textit{'West South'} or $V(T_{West}) = \{\textit{'West'}, \textit{'West North'}, \textit{'West South'}\}$.

\subsubsection{Time-steps}

\begin{figure}
	\centering
		\includegraphics[scale=.5]{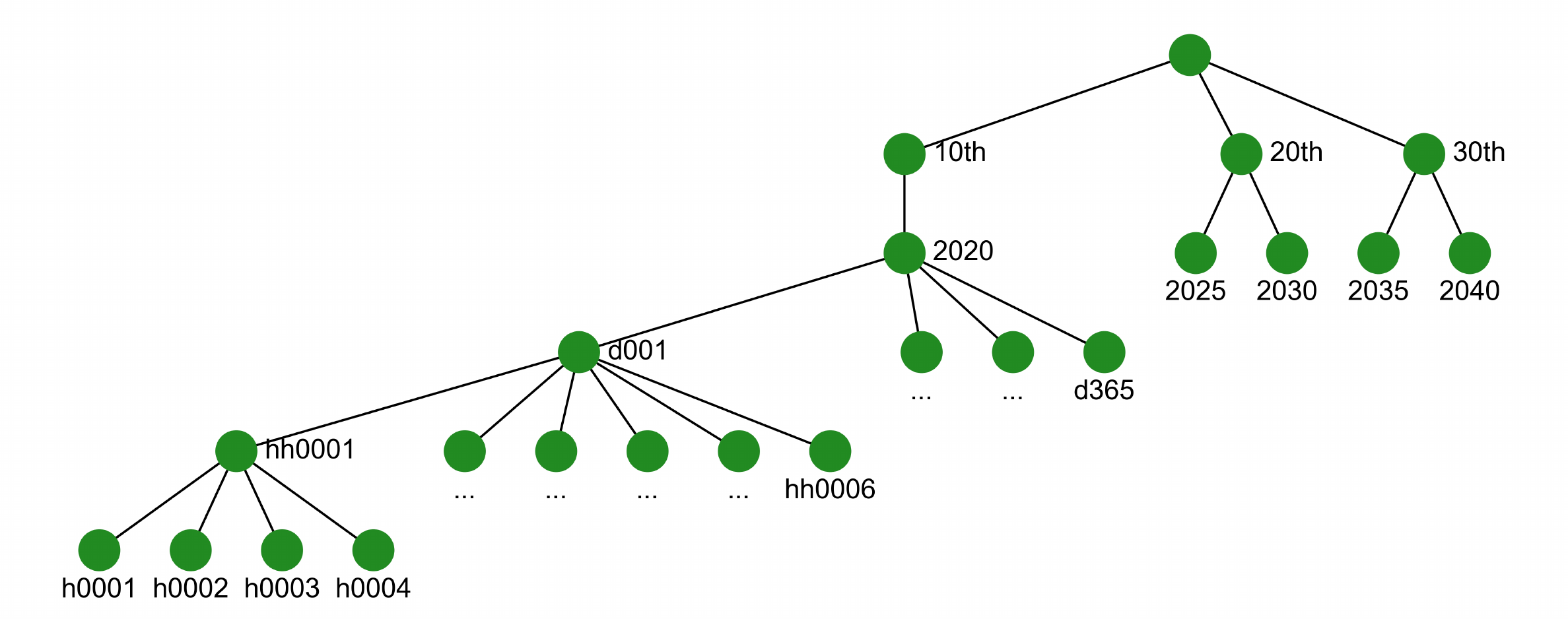}
	\caption{Rooted tree of time-steps in the example model}
	\label{fig:2}
\end{figure}

Analogously to regions, time-steps are organized in the rooted tree $T$ with $t$ representing an arbitrary vertex. In a reduced form, for the example model this tree is drawn in Fig. \ref{fig:2}. Vertices with depth one each represent a decade, vertices with depth two correspond to all years considered within the respective decade and each year is then further dissected into daily, four-hour, and finally hourly steps.

\subsubsection{Carriers} \label{23}

\begin{figure}
	\centering
		\includegraphics[scale=.5]{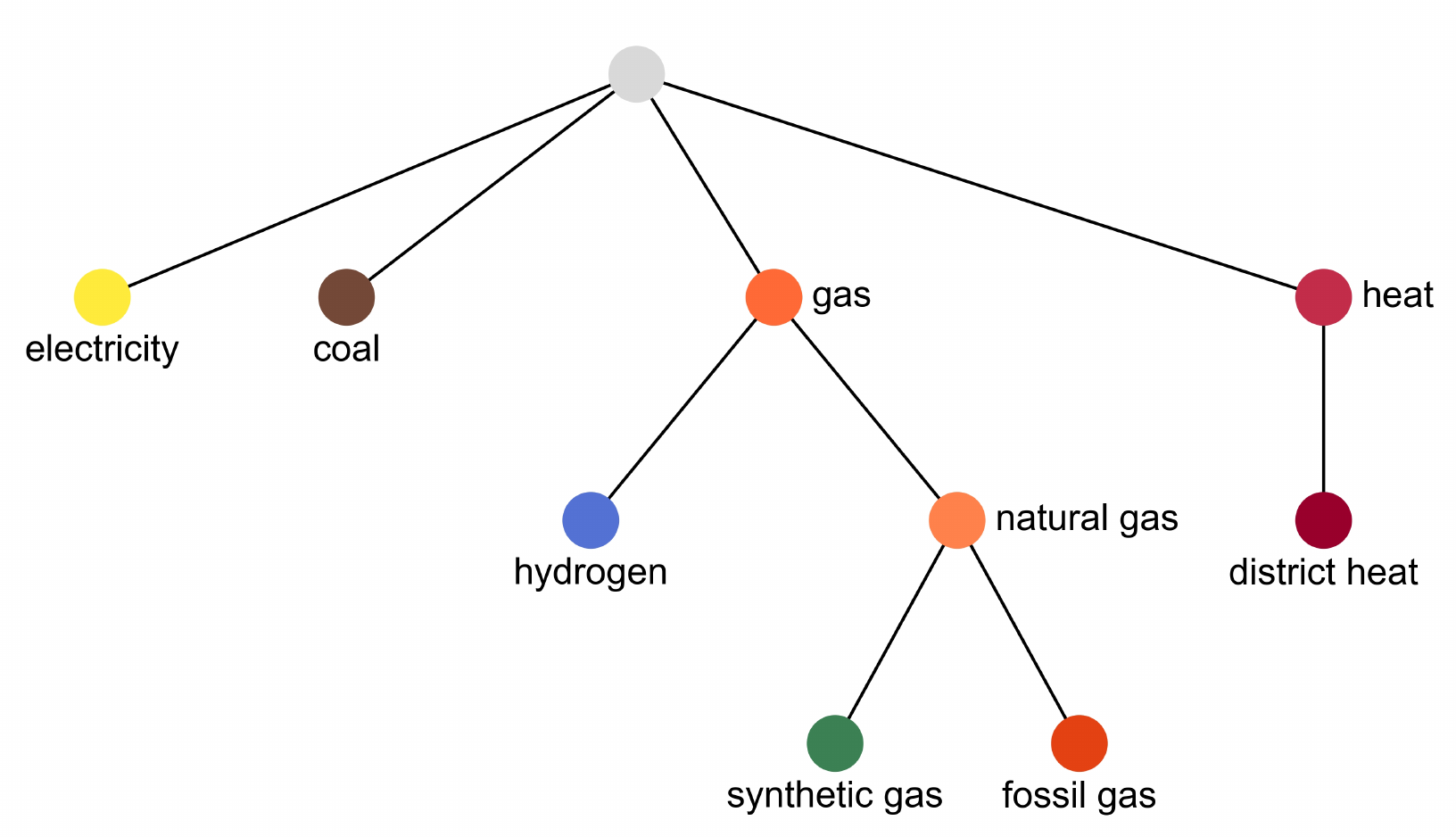}
	\caption{Rooted tree of energy carriers in the example model}
	\label{fig:3}
\end{figure}

Fig. \ref{fig:3} displays the rooted tree $C$ of all energy carriers defined within the model. While the vertices \textit{'coal'} and \textit{'electricity'} do not have any descendants, \textit{'heat'}, which only refers to low-temperature heat, has one descendant \textit{'district heat'} and gases are subdivided into \textit{'hydrogen'} and \textit{'natural gas'}, which again is split into \textit{'synthetic gas'} and \textit{'fossil gas'}. This arrangement is motivated by the fact that having carriers share a common ancestor is required for modeling them as substitutes in a certain context, as we will elaborate in section \ref{3}.

To specify the temporal and spatial granularities carriers are modeled at, each are assigned depths within the rooted trees of time-steps and regions. This is done separately for dispatch and expansion and summarized for the example model in Tab. \ref{table:1}.
\begin{table}[width=.9\linewidth,cols=6,pos=h]
\caption{Depths assigned to energy carriers in the example model}\label{tbl1}
\begin{tabular*}{\tblwidth}{@{} LLLCCCC@{} }
\toprule
\multirow{2}{*}{carrier with depth 1} & \multirow{2}{*}{carrier with depth 2} & \multirow{2}{*}{carrier with depth 3} & \multicolumn{2}{c}{temporal} & \multicolumn{2}{c}{spatial} \\
\cline{4-5}\cline{6-7}
& & & \text{\footnotesize dispatch}  & \text{\footnotesize expansion} & \text{\footnotesize dispatch} & \text{\footnotesize expansion} \\
\midrule
electricity &  & & 5 & 2 & 1 & 1 \\
heat &  district heat & & 4 & 2 & 2 & 2 \\
gas & natural gas & synthetic gas & 3 & 2 & 1 & 1 \\
gas & natural gas & fossil gas & 3 & 2 & 1 & 1 \\
gas & hydrogen & & 3 & 2 & 1 & 1 \\
coal & & & 2 & 2 & 1 & 1\\
\bottomrule
\end{tabular*}
\label{table:1}
\end{table}

Consequently, a depth of five for temporal dispatch of \textit{'electricity'} means dispatch of the carrier is modeled for every time-step with depth five, which, going back to Fig. \ref{fig:2}, corresponds to an hourly granularity. Likewise, \textit{'heat'} and \textit{'district heat'} are modeled at four-hour steps and all gases are balanced daily. Lastly, \textit{'coal'} is only accounted for per year. Deciding on the temporal granularity of dispatch for a carrier is a crucial assumption on its inherent flexibility. For electricity an hourly resolution is often considered adequate when using spatially aggregated models \citep{Brown2018b}. As a result of its physical properties, gas, in contrast to electricity, is traded daily. In accordance with dedicated literature, a daily resolution is also applied here \citep{Hauser2019,Petrovic2017}. For heat, a four-hour resolution was assumed to account for the thermal inertia of buildings. 

The uniform depth of two for all carriers' temporal expansion granularity means decisions on capacity expansion are made for each year. If the depth were set to one instead, a decision on expansion would apply for an entire decade. Such a setup would be suited to mimic typical polices for the expansion of wind and solar capacities. 

Spatial dispatch and expansion granularity for all carriers corresponds to the regions with depth 1, namely \textit{'West'} and \textit{'East'}, except for \textit{'heat'} and \textit{'district heat'}. Here a more detailed resolution was chosen, since heat, unlike electricity or gas, cannot be transported over greater distances to offset local imbalances between supply and demand.

Certain conditions can be defined that ensure the temporal and spatial granularities assigned to each carrier are suited to create a logical consistent energy system model. AnyMOD specifically checks compliance of these conditions and throws an error, if any of them is violated. To formulate these rules, the depths mapped to a specific carrier $c$ will be termed $dep_{c}$.

First, a carrier may not be modeled at a dispatch granularity more detailed than any of its descendants, regardless if temporal or spatial. This means, the depth assigned to a specific carrier cannot exceed the smallest depth assigned to any of its descendants, as denoted in Eqs. \ref{eq:1} and \ref{eq:2}.\footnote{The hat operator is used throughout the paper to indicate a vertex is a descendant to another vertex within the same equation.}
\begin{subequations}  
\begin{align} 
    dep_{c}^{dis,tp} \leq \min_{\hat{c} \in \delta_{c}}{dep_{\hat{c}}^{dis,tp}} && \forall c \in V(C) \label{eq:1} \\
    dep_{c}^{dis,sp} \leq \min_{\hat{c} \in \delta_{c}}{dep_{\hat{c}}^{dis,sp}} && \forall c \in V(C) \label{eq:2}
\end{align}
\end{subequations} 
The conditions originate from the way the framework models substitution of energy carriers. As section \ref{3} will explain in detail, this is achieved by aggregating variables of descendant carriers with the ancestral carrier. However, such an aggregation is impossible, if for example the ancestral carrier has an hourly resolution, but one of its descendants is modeled daily.

The second group of conditions addresses the relation between dispatch and expansion granularity. As stated in Eq. \ref{eq:3}, the spatial granularity of expansion may not be less detailed than the spatial granularity of dispatch for any carrier or, in terms of depths, the depth of dispatch cannot exceed the depth of expansion.
\begin{align}
    dep_{c}^{exp,sp} \geq dep_{c}^{dis,sp} && \forall c \in V(C) \label{eq:3} 
\end{align}
This condition is necessary to ensure each dispatch variable in the model can be mapped to a corresponding capacity. If, for instance, expansion is modeled at the country level, but dispatch considered separately for each state within the country, assigning a capacity to each of these states would not be possible. The opposite case with dispatch on the country level but regional expansion is supported and leads to an aggregation of regional capacities by country.

For the same reason a similar condition on temporal granularities is required. This condition states that for any carrier the temporal granularity of expansion may not be more detailed than the temporal granularity of dispatch. As formulated in Eq. \ref{eq:4}, this implies the depth assigned for expansion cannot exceed the depth of dispatch.
\begin{align}
 dep_{c}^{exp,tp} \leq dep_{c}^{dis,tp} && \forall c \in V(C) \label{eq:4} 
\end{align}
If, in violation of Eq. \ref{eq:4}, capacity expansion had an daily resolution, but dispatch were only modeled yearly, again a sensible assignment of capacity to dispatch variables would not be possible.

Modeling several periods of capacity expansion requires to define superordinate dispatch time-steps. Dispatch within each of these steps is self-contained, meaning dispatch decisions within the period do not affect any of the other periods. For instance, cyclic conditions for storage will enforce the same storage levels at the beginning and end of each of those periods. This also implies that capacities cannot vary within these periods. Most existing models take a yearly resolution for this purpose, but other granularities are conceivable as well.\footnote{Even varying this resolution within the model is theoretically possible, but does not appear practical and was not implemented.} Since these periods connect expansion and dispatch, their depth, denoted as $dep^{sup}$, must be within the interval from the most detailed expansion resolution to the least detailed dispatch resolution. This is expressed by Eq. \ref{eq:5}:
\begin{gather}
dep^{sup} \in [  \max_{c \in V(C)} dep_{c}^{exp,tp}, \, \min_{c \in V(C)} dep_{c}^{dis,tp}] \label{eq:5} 
\end{gather}
$\Phi$ is defined as the set of all superordinate dispatch time-steps. Each subordinate dispatch time-step $t$ has exactly one ancestor within $\Phi$, which is referred to as $\alpha_{t}^{sup}$. In the example model $dep^{sup}$ is two and consequently $\Phi$ corresponds to all years. For any hour or day $t$, $\alpha_{t}^{sup}$ assigns the year the respective day or hour is in.

\subsubsection{Technologies} \label{24}

\begin{figure}
	\centering
		\includegraphics[scale=.55]{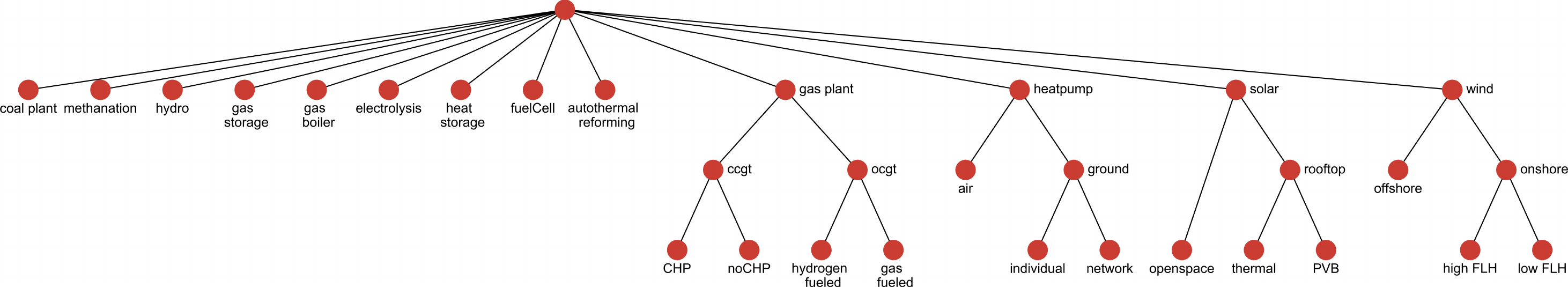}
	\caption{Rooted tree of technologies in the example model}
	\label{fig:4}
\end{figure}

Technologies are organized in the rooted tree $E$, which is shown in Fig. \ref{fig:4} for the example model. Only leaves of this tree correspond to actual technologies, while all other vertices serve the sole purpose of organizing them. For instance, to reflect how photovoltaic and solar thermal rooftop systems compete for a limited amount of rooftop area, their shared ancestor \textit{'rooftop'} can be used to enforce an upper limit on the sum of their capacities.

The function $g(e)$ maps technologies to one of three groups: \textit{stock}, \textit{mature} and \textit{emerging}. Stock technologies cannot be expended and are limited to pre-existing capacities. Emerging technologies differ from mature technologies in the sense that their capacities are differentiated by time-step of construction. In the case of electrolyzers for example, substantial increases in efficiency are expected by 2050. To account for such improvements, capacities build in different years have to be considered separately. For mature technologies, no substantial advances are expected, and such differentiation would only cause an unnecessary increase in model size.

Generated and used carriers are mapped to technologies by the sets $\gamma_{e}^{gen}$ and $\gamma_{e}^{use}$, respectively. Any used carrier $c$ cannot be a descendant to another used carrier $c'$. The condition applies to generated carriers analogously and both conditions are formalized by Eqs. \ref{eq:7} and \ref{eq:8}.
\begin{subequations}  
\begin{gather}
c \notin \delta_{c'} \; \; \; \forall e \in V(E), (c,c') \in \{ \gamma_{e}^{use} \, \times \, \gamma_{e}^{use} \, | \, c \neq c' \} \label{eq:7} \\
c \notin \delta_{c'} \; \; \; \forall e \in V(E), (c,c') \in \{ \gamma_{e}^{gen} \, \times \, \gamma_{e}^{gen} \, | \, c \neq c' \} \label{eq:8} 
\end{gather}
\end{subequations}
Considering the combined-cycle gas power plant (CCGT) with cogeneration (CHP) from the example, \textit{'natural gas'} is converted to \textit{'district heat'} and \textit{'electricity'}, hence $\gamma_{CHP}^{use} =  \{\textit{'natural gas'}\}$ and $\gamma_{CHP}^{gen} =  \{\textit{'district heat'},\textit{'electricity'}\}$. Additionally assigning \textit{'fossil gas'} as a used carrier would pose a logical contradiction since \textit{'natural gas'} implicitly already includes its descendant \textit{'fossil gas'} and consequently violate Eq. \ref{eq:7}.

Charged carriers are denoted as $\gamma_{e}^{stI}$; discharged carriers are referred to as $\gamma_{e}^{stO}$. By default, only carriers, which are leaves, can be explicitly stored. If a technology is defined to store a non-leaf carrier $c$, actually stored are only its leaves $\lambda_{c}$. For instance, in the example \textit{gas storage} is defined to store \textit{gas} which means the technology can equally store \textit{hydrogen}, \textit{synthetic gas} and \textit{fossil gas}. Deviating from this approach gives rise to unintended effects.\footnote{It can be explicitly enforced though, but this a special case not discussed within the paper.} To elucidate this, assume \textit{gas storage} would directly store the carrier \textit{gas} instead. Since descendants are included in the ancestors energy balance, \textit{hydrogen} could still be charged. However, it would be discharged as \textit{gas} and could not be used wherever \textit{hydrogen} is specifically required.

The representation of storage is not limited to charging and discharging carriers from external sources but can also account for carriers generated or used by the same technology. To clarify this, we assume a carrier $c$ is an element of $\gamma_{e}^{stO}$, but not within $\gamma_{e}^{stI}$. This implies it can be discharged, but not charged from an external source. However, if $c$ is also an element of $\gamma_{e}^{gen}$, it can be charged by the technologies own generation instead. For instance, the photovoltaic battery system (PVB) in the example represents a photovoltaic panel combined with a home battery. In line with other research, we assume home batteries cannot be charged from the grid, but can provide electricity to the grid \citep{Schopfer2018}. Therefore, \textit{'electricity'} is an element of $\gamma_{PVB}^{gen}$ and $\gamma_{PVB}^{stO}$, but $\gamma_{PVB}^{stI}$ is empty. Nevertheless, the battery can still be charged by the system's own generation from the photovoltaic panel. Correspondingly, a charged carrier can be discharged internally if within $\gamma_{e}^{use}$. In this case, an industrial furnace provided with gas by an on-site gas storage could serve as an example. If carriers are charged or discharged internally, also non-leaf carriers can be stored.

Applying this, \ref{eq:108} and \ref{eq:9} define sets of stored carriers for a technology $e$. All carriers charged and discharged externally are provided by $\gamma_{e}^{stEx}$. This set is unified with all carriers charged externally and discharged internally as well as the other way around, to obtain all carriers stored $\gamma_{e}^{st}$.
\begin{gather}
\gamma_{e}^{stEx} \coloneqq \overbrace{\{\lambda_{c} : c \in \gamma_{e}^{stO} \cup \gamma_{e}^{stI}\}}^ \text{external charging or discharging} \label{eq:108} \\
\gamma_{e}^{st} \coloneqq \gamma_{e}^{stEx} \cup \underbrace{(\gamma_{e}^{gen} \cup \gamma_{e}^{stO})}_\text{internal charging} \cap \underbrace{(\gamma_{e}^{use} \cap \gamma_{e}^{stI})}_\text{internal discharging} \label{eq:9}
\end{gather}
The sets $\gamma_{e}^{in}$ and $\gamma_{e}^{out}$ collect all external in- and output carriers of a technology $e$:
\begin{subequations}
\begin{gather}
\gamma_{e}^{in} \coloneqq \gamma_{e}^{use} \cup \gamma_{e}^{stEx} \\
\gamma_{e}^{out} \coloneqq \gamma_{e}^{gen} \cup \gamma_{e}^{stEx}
\end{gather}
\end{subequations}
In addition, all technologies any conversion or storage carrier was assigned to are collected within the respective sets $\Gamma^{cv}$ and $\Gamma^{st}$, which are defined by the following equations:
\begin{subequations}
\begin{gather}
\Gamma^{cv} \coloneqq \{V(E) \, | \, \gamma_{e}^{gen} \cup \gamma_{e}^{use} \neq \emptyset \} \\
\Gamma^{st} \coloneqq \{V(E) \, | \, \gamma_{e}^{st} \neq \emptyset \}
\end{gather}
\end{subequations}

\begin{figure}
	\centering
		\includegraphics[scale=.5]{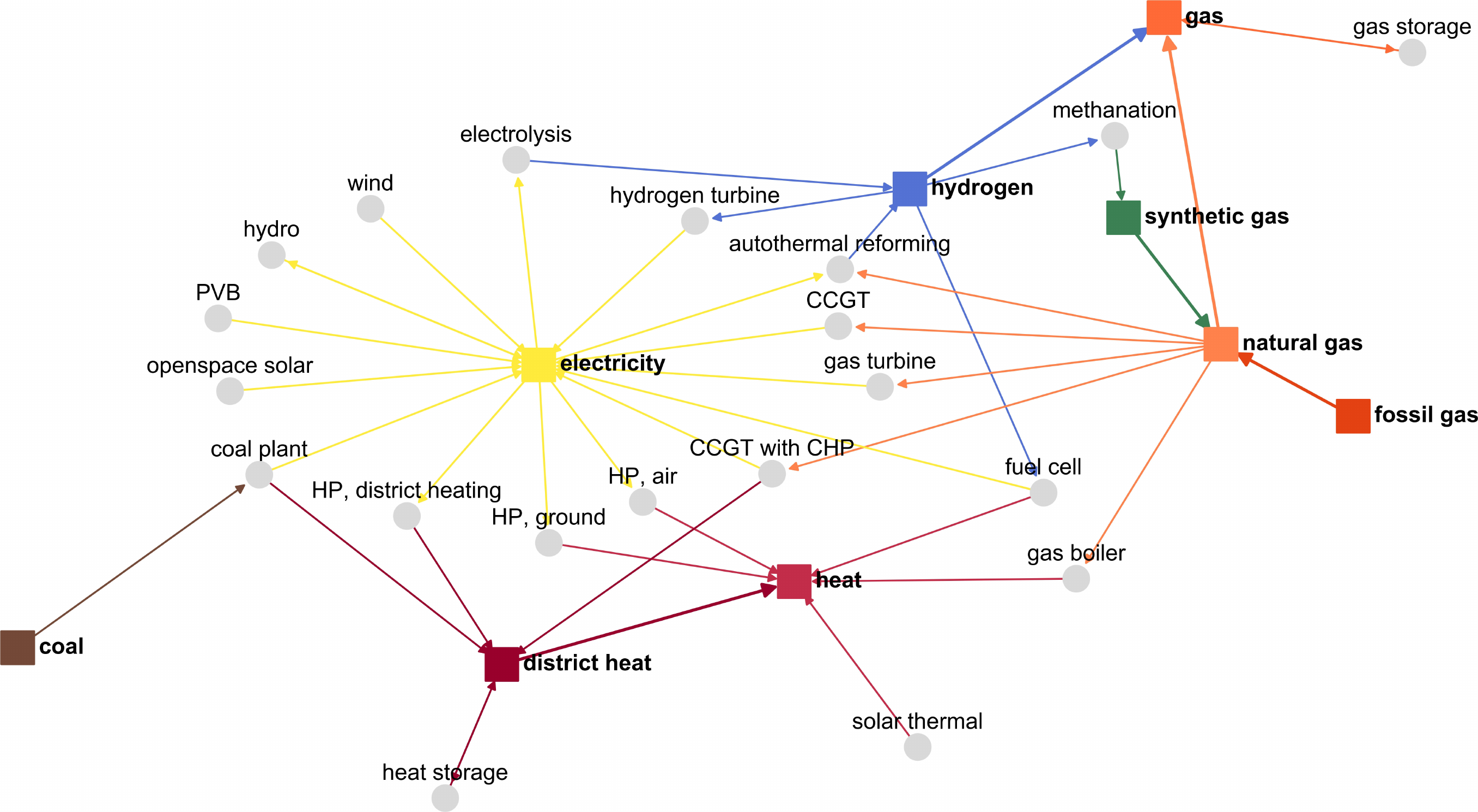}
	\caption{Qualitative energy flow graph for example model}
	\label{fig:5}
\end{figure}

The directed graph in Fig. \ref{fig:5} summarizes how in- and output carriers are mapped to technologies in the example model. In the graph all technologies are symbolized by grey vertices. Their entering edges relate to inputs $\gamma_{e}^{in}$; outgoing edges to outputs $\gamma_{e}^{out}$. Carriers are symbolized by colored vertices that have outgoing edges directed towards their ancestors. The graph demonstrates, how organizing carriers in rooted trees supports modeling the manifold ways energy carriers can be substituted and interact with technologies in an integrated energy system: Synthetic gas must be created from hydrogen, which again requires the use of electricity via electrolysis, while natural gas cannot be created from other carriers. However, both energy carriers can equally fuel gas boilers and power plants or be used for auto thermal reforming, a gas-based process to create hydrogen. Also, any of these carriers can be stored in a gas storage system, since \textit{gas} is an ancestor to all of them. 

\begin{figure}
	\centering
		\includegraphics[scale=.5]{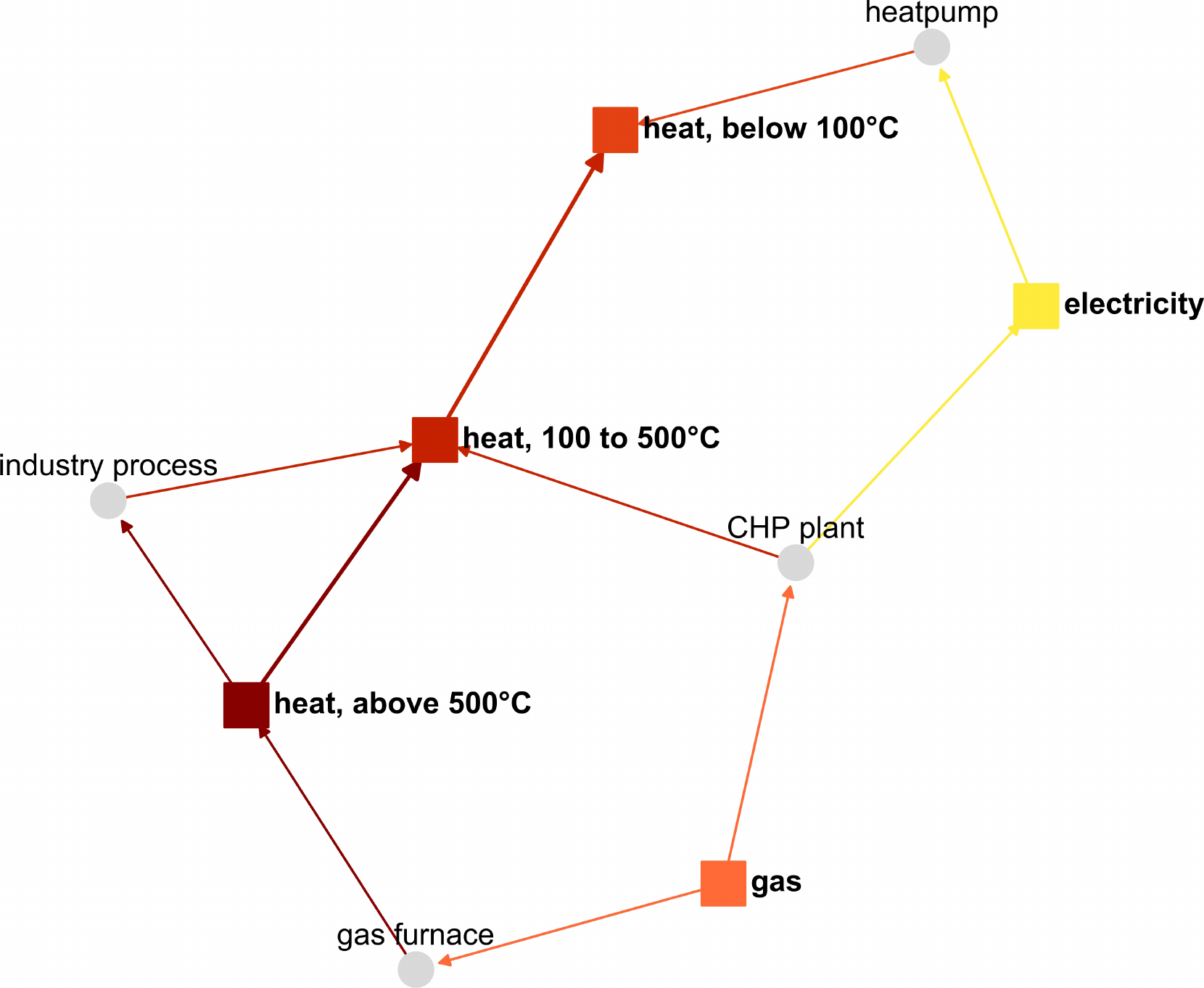}
	\caption{Qualitative energy flow graph for alternative application}
	\label{fig:6}
\end{figure}

Although the example model focuses on the interplay of gas-based fuels to demonstrate the capabilities of the presented method, it can be applied beyond: For instance, processes in the energy-intensive industry often require high-temperature heat at different levels, which makes decarbonization challenging \citep{bataille2018}. However, providing a process with heat on a temperature level that exceeds its requirements is possible. Also, excess heat from one process can serve as an input to another. The qualitative energy flow diagram in Fig. \ref{fig:6} outlines how these aspects could be accounted for within energy system models by the introduced method. Since the carrier \textit{'heat, above 500°C'} is a descendant of \textit{'heat, 100 to 500°C'} and \textit{'heat, below 100°C'}, in contrast to the other technologies \textit{'gas furnace'} is able to satisfy demand on all levels. Also, a process that requires heat at the highest temperature level and provides excess heat again at a lower level, can be modeled.

\subsubsection{Modes}

The rooted tree $M$ organizes the different operational modes $m$ defined within the framework. In contrast to the other graphs, the rooted tree of modes is trivial, meaning it only consists out of the root and its direct descendants. The set $\mu_{e}$ maps its operational modes to each technology or, if only one mode exists, just assigns the root $rt_{M}$. In the example model, distinct modes termed \textit{more heat} and \textit{more electricity} are only defined for CCGT plants with CHP. The \textit{more Heat} mode operates at a higher fuel utilization rate, but a smaller CHP coefficient.

\subsection{Equations of optimization problem} \label{3}

Building on these sets and mappings, the constraints of the model's underlying optimization can be formulated. We start with dispatch related constraints, followed by capacity constraints, which connect dispatch and expansion and close with the equations to describe expansion. Since the cost minimizing objective function does not substantially differ to pre-existing formulations, it is provided in Appendix B. The same applies for constraints that impose exogenous limits on variables.
\subsubsection{Energy balance}

The energy balance ensures demand for each carrier $c$ equals or does not exceed its supply at any time $t$ or place $r$. To model this, all dispatch time-steps $\tau_{c}$ and regions $\rho_{c}$ of a carrier $c$ are defined as follows:
\begin{subequations}
\begin{gather}
\tau_{c} \coloneqq  \{ V(T) \, | \, d(t) = dep_{c}^{dis,tp}\} \label{eq:10} \\
\rho_{c} \coloneqq  \{ V(R) \, | \, d(r) = dep_{c}^{dis,sp}\} \label{eq:11} 
\end{gather}
\end{subequations}
Consequently, the cartesian product of $\tau_{c}$ and $\rho_{c}$ gives the temporal and spatial granularity $\varphi_{c}$ that a carrier $c$ is modeled at.
\begin{gather} 
\varphi_{c} \coloneqq \tau_{c} \times \rho_{c}
\end{gather}

Since demand for a carrier $c$ can not only be met by the carrier itself, but also by its descendants $\hat{c}$, these have to be included into the energy balance as well. However, according to Eq. \ref{eq:1} and \ref{eq:2}, these descendants might be modeled at a granularity more detailed than the carrier itself.
Therefore, elements of these descendants must be aggregated to comply with the resolution of the ancestral carrier. When balancing the time-step $t$, the dispatch time-steps of a descendant carrier $\hat{c}$ that require aggregation, correspond to the intersection of descendant carriers time-steps $\tau_{\hat{c}}$ with the descendants of the balanced time-step $\delta_{t}^+$. The same reasoning is applied to regions and the set of pairs $\sigma_{\hat{c},r,t}$ can be obtained. As defined by Eq. \ref{eq:12}, this set contains all time-steps and regions that have to be aggregated to account for dispatch of a carrier $\hat{c}$ at time-step $t$ in region $r$.
\begin{gather}
    \sigma_{\hat{c},r,t} \coloneqq \tau_{\hat{c}} \cap \delta_{t}^+ \, \times \,  \rho_{\hat{c}} \cap \delta_{r}^+ \label{eq:12} 
\end{gather}
The equation applies as well, if $t$ or $r$ are already at the right granularity, because the set $\delta_v^+$ by definition also includes the vertex $v$ itself. 

To enable descendant carriers to satisfy demand, by default the energy balance is not an equality constraint and supply might exceed demand. The carriers \textit{district heat} and \textit{heat} from the example can be used to illustrate this. To let the model endogenously decide whether to use district heating technologies or not, demand was only specified for the ancestral carrier \textit{heat}.\footnote{In the example, an upper limit on the generation of district heat for each time-step reflects that only a share of consumers can be connected to a district heating network.} As a result, formulating the energy balance for \textit{district heat} as an equality constraint, would fix its generation to zero. 

Building on this, the energy balance is formulated in eqn. \ref{eq:13}. To facilitate the understanding, optimization variables have capital initials, while parameters are written in lowercase.

\begin{align}
\sum_{\hat{c} \in \delta_{c}^{+}} \; \sum_{\langle \hat{t},\hat{r} \rangle \in \sigma_{t,r,\hat{c}}}  \underbrace{Te_{\hat{t},\hat{r},\hat{c}}^{cv} + Te_{\hat{t},\hat{r},\hat{c}}^{st}}_\text{\makebox[0pt][c]{supply and demand by technologies}} \; + \; \overbrace{Exc_{\hat{t},\hat{r},\hat{c}}^{net}}^\text{\makebox[0pt][c]{exchange with other regions}} \; + \; \underbrace{Trd_{\hat{t},\hat{r},\hat{c}}^{net}}_\text{\makebox[0pt][c]{trade with exogenous markets}} \; - \; \overbrace{dem_{\hat{t},\hat{r},\hat{c}}^{\text{\makebox(10,4.6)[c]{\,}}}}^\text{\makebox[0pt][c]{exogenous demand}} \geq 0 &&	\forall c \in V(C), \, \langle t, r \rangle \in \varphi_{c} \label{eq:13} 
\end{align}

Conversion related dispatch variables are summarized by $Te^{cv}$ and include $Gen$ for generation and $Use$ for use. Analogously, $Te^{st}$ is composed of $StI^{ext}$ and $StO^{ext}$ to account for external in- and output of storage. Each of these variables is specified for five different dimensions: time-step of dispatch $t$, region $r$, carrier $c$, mode $m$ and lastly time-step of construction $\tilde{t}$. The cartesian product of all dimensions is denoted as $\Omega$.

For stock and mature technologies, which are not differentiated by time-step of construction, $\tilde{t}$ always corresponds to the root of the time-step tree $rt_{T}$. In case of an emerging technology, all time-steps of construction that result in a lifespan, which includes the dispatch time-step $t$, have to be considered separately. To elucidate this, consider the an emerging technology with a constant lifetime $lt_{e,\tilde{t}}$ of 15 years. For any dispatch time-step $t$ within the year 2020, only capacities constructed in 2020 have to be considered. However, if $t$ is within 2050 instead, the construction time-steps 2040 and 2045 have to considered in addition to 2050. In conclusion, Eq. \ref{eq:14} defines the set $\theta_{e,t}^{dis}$ that provides the construction time-steps to consider separately for a technology $e$ at dispatch time-step $t$.
\begin{align}
\theta_{e,\tilde{t}}^{dis} \coloneqq \begin{cases} 
      \{r_{T} \} & \text{,if } g(e) = \textit{`mature'} \lor g(e) = \textit{`stock'} \\
      \{\tilde{t}' \in \Phi \, | \, \tilde{t}' \in (\alpha_{t}^{sup} -lt_{e,\tilde{t}'} ,\alpha_{t}^{sup}] \} & \text{,if } g(e) = \textit{`emerging'} 
   \end{cases} \label{eq:14} 
\end{align}

Dispatch variables for all conversion and storage technologies are summed by time-step of construction $\tilde{t}$ and modes $m$ to define $Te^{cv}$ and $Te^{st}$  as denoted in Eqs. \ref{eq:17} and \ref{eq:18}. Iverson brackets are used to indicate that dispatch variables are only created, if the respective carrier is actually assigned to the technology.
\begin{subequations}
\begin{align}
Te_{t,r,c}^{cv} =  \sum_{e \in \Gamma^{cv}} \; \sum_{\tilde{t} \in \theta_{e,t}^{dis}} \; \sum_{m \in \mu_{e}} Gen_{t,\tilde{t},r,c,e,m}[c \in \gamma_{e}^{gen}] - Use_{t,\tilde{t},r,c,e,m}[c \in \gamma_{e}^{use}] 
               && \forall c \in V(C), \, \langle t, r \rangle \in \varphi_{c}  \label{eq:17} \\
Te_{t,r,c}^{st} =  \sum_{e \in \Gamma^{st}} \; \sum_{\tilde{t} \in \theta_{e,t}^{dis}} \; \sum_{m \in \mu_{e}} StO_{t,\tilde{t},r,c,e,m}^{ext}[c \in \gamma_{e}^{stEx}] - StI_{t,\tilde{t},r,c,e,m}^{ext}[c \in \gamma_{e}^{stEx}] && \forall c \in V(C), \, \langle t, r \rangle \in \varphi_{c} \label{eq:18} 
\end{align}
\end{subequations}

In the energy balance, $Exc_{\hat{t},\hat{r},\hat{c}}^{net}$ refers to net imports of region $\hat{r}$ from other regions. The set $\beta_{c,r}$ includes all regions with that region $r$ can exchange carrier $c$. Exchange can be considered similar to storage, since both shift energy, one in space and the other in time. Therefore, exchange of carriers is limited to leaves, because otherwise the same effects as described for storage earlier will occur. For instance, to represent the gas network in the example, $\beta_{c,r}$ is defined for \textit{gas}. Consequently, only the carriers \textit{hydrogen}, \textit{synthetic gas} and \textit{fossil gas} are explicitly exchanged.

Applying this, Eq. \ref{eq:19} computes the net import based on the exchange variables $Exc$ and the efficiency of exchange $eff^{exc}$ that accounts for exchange losses. The first region in the index always refers to the region energy is being transported to and the second to the region it is being transported from.
\begin{equation}
Exc_{t,r,\hat{c}}^{net} = \sum_{r' \in \beta_{c,r}} \; \sum_{\hat{r}' \in \rho_{\hat{c}} \cap \delta_{r'}^+ }  \frac{Exc_{t,\hat{r},\hat{r}' ,c}}{1/eff_{t,\hat{r},\hat{r}' ,\hat{c}}^{exc}}  - Exc_{t,\hat{r}' ,\hat{r},\hat{c}} \;
\forall \langle c,r \rangle \in \{V(C) \times V(R) \, | \, \beta_{c,r} \}, \, \hat{c} \in \lambda_{c}, \, t \in \tau_{\hat{c}}, \, \hat{r} \in  \rho_{\hat{c}} \cap \delta_{r}^+ \label{eq:19} 
\end{equation}
Just as explained at the beginning of the section, the region specified in $\beta_{c,r}$ might be less detailed than the regions a descendant carrier $\hat{c}$ is modeled for. Therefore, exchange variables are aggregated by regions using the same formulation introduced earlier.

The net effect of trade is accounted for in the energy balance by $Trd^{net}$ defined in Eq. \ref{eq:20}. In contrast to exchange, trade refers to buying or selling carriers to an exogenous market at a fixed price. The quantity that can be bought or sold at a given price can be limited, which can be used to create a stepped supply or demand curve. Each of these steps is denoted as $\zeta^{buy}$ or $\zeta^{sell}$, respectively.
\begin{align}
Trd_{t,r,c}^{net} = \sum_{i \in \zeta^{buy}} Trd_{t,r,c,i}^{buy} - \sum_{i \in \zeta^{sell}} Trd_{t,r,c,i}^{sell} && \forall c \in V(C), \, \langle t, r \rangle \in \varphi_{c} \label{eq:20} 
\end{align}
Potential applications of this functionality range from a representation of commodity markets to accounting for price-elastic demand in the electricity sector. The last remaining element of the energy balance $dem$ is an exogenously set parameter and refers to inelastic demand.

\subsubsection{Conversion balance}

The conversion balance describes how technologies transform energy carriers into one another. For this purpose, the in- and outputs to the conversion process are summarized by carrier as $Cv^{in}$ and $Cv^{out}$, which are defined in Eqs. \ref{eq:21} and \ref{eq:22}. As set out in section \ref{24}, these in- and outputs are not limited to use and generation variables, but can also include internal storage variables.
\begin{subequations} 
\begin{align}
Cv_{t,\tilde{t},r,c,e,m}^{in} = Use_{t,\tilde{t},r,c,e,m} + StO_{t,\tilde{t},r,c,e,m}^{int}[c \in \gamma_{e}^{stO}] && \forall e \in V(E), \,c \in \gamma_{e}^{use}, \langle t, \,r \rangle \in \varphi_{c}, \,\tilde{t} \in \theta_{e,t}^{dis},  \,m \in \mu_{e} \label{eq:21} \\
Cv_{t,\tilde{t},r,c,e,m}^{out} = Gen_{t,\tilde{t},r,c,e,m} + StI_{t,\tilde{t},r,c,e,m}^{int}[c \in \gamma_{e}^{stI}] && \forall e \in V(E), \,c \in \gamma_{e}^{gen}, \, \langle t, r \rangle \in \varphi_{c}, \,\tilde{t} \in \theta_{e,t}^{dis}, \,m \in \mu_{e}  \label{eq:22}
\end{align}
\end{subequations}

Only technologies that are assigned both, used and generated carriers, require a conversion balance. Conversion is balanced at the least detailed granularity of all carriers involved. Otherwise, a carrier with a less detailed granularity could not be accounted for. Applying this, Eq. \ref{eq:23} defines the resolution of the energy balance for each technology $e$.
\begin{equation}
\epsilon_{e} \coloneqq \{ V(T) \, | \, d(t) =  \min\limits_{c \in \gamma_{e}^{gen} \cup \gamma_{e}^{use}}dep_{c}^{dis,tp} \} \times \{ V(R) \, | \, d(r) = \min\limits_{c \in \gamma_{e}^{gen} \cup \gamma_{e}^{use}}dep_{c}^{dis,sp}\} \label{eq:23} 
\end{equation} 
The overall efficiency of a conversion process that determines the ratio between in- and output quantities is denoted as $eff^{cv}$. If a technology's conversion efficiency differs by operational mode, each of these modes must be considered by a separate equation. Therefore, $\omega^{cv}$ provides all sets of modes that require an individual balance. On this basis, the conversion balance given by Eq. \ref{eq:24} can be formed.
\begin{equation}
\begin{split}
\sum_{m \in \xi} eff_{t,\tilde{t},r,e,m}^{cv} \sum_{c \in \gamma_{e}^{use}} \; \sum_{\langle \hat{t},\hat{r} \rangle \in \sigma_{t,r,c}} Cv_{\hat{t},\tilde{t},\hat{r},c,e,m}^{in}  = \sum_{m \in \xi}  \; \sum_{c \in \gamma_{e}^{gen}} \; \sum_{\langle \hat{t},\hat{r} \rangle \in \sigma_{t,r,c}} Cv_{\hat{t},\tilde{t},\hat{r},c,e,m}^{out} \\&\hspace{-140pt} \forall e \in \{V(E) \, | \, \gamma_{e}^{use} \neq \emptyset  \land  \gamma_{e}^{gen} \neq \emptyset\}, \, \langle t,r \rangle \in \epsilon_{e}, \tilde{t} \in \theta_{e,t}^{dis}, \, \xi \in \omega_{t,\tilde{t},r,e}^{cv} \label{eq:24}
\end{split}
\end{equation}
For the CCGT plant with CHP from the example, the conversion balance is created daily and for each region of depth one, which corresponds to the granularity of its least detailed carrier \textit{gas}. In addition, separate balances are created for each operational mode, since these differ in terms of efficiency, which means $\omega^{cv} = \{\{\textit{'more heat'}\},\{\textit{'more electricity'}\} \}$.

\subsubsection{Storage balance} \label{33}

The storage balance connects in- and output of a storage system to the storage level. The in- and output to the storage are comprised of external and internal storage variables as defined in Eq. \ref{eq:26}.
\begin{subequations} \label{eq:26}
\begin{align}
St_{t,\tilde{t},r,c,e,m}^{in} = StI_{t,\tilde{t},r,c,e,m}^{ext} + StI_{t,\tilde{t},r,c,e,m}^{int} && \forall e \in V(E), \,c \in \gamma_{e}^{st}, \,\langle t, r \rangle \in \varphi_{c}, \,\tilde{t} \in \theta_{e,t}^{dis}, \,m \in \mu_{e}  \\
St_{t,\tilde{t},r,c,e,m}^{out} = StO_{t,\tilde{t},r,c,e,m}^{ext} + StO_{t,\tilde{t},r,c,e,m}^{int} && \forall e \in V(E), \,c \in \gamma_{e}^{st}, \,\langle t, r \rangle \in \varphi_{c}, \,\tilde{t} \in \theta_{e,t}^{dis}, \,m \in \mu_{e} 
\end{align}
\end{subequations} 
In Eq. \ref{eq:27} the storage level $StLvl$ at time-step $t$ is computed by summing levels of the previous time-step $t-1$ with storage in- and outputs. To a enforce a cyclic condition, the previous time-step to the first time-step is the last time-step within the same superordinate dispatch time-step (i.e. for \textit{h0001} in \textit{2020} the previous time-step is \textit{h8760} in \textit{2020}).
\begin{equation}
\begin{split}
\overbrace{\sum_{m \in \xi} StLvl_{t,\tilde{t},r,c,e,m}^{\text{\makebox(10,15)[c]{\,}}}}^\text{current level}  = \overbrace{\sum_{m \in \xi} \frac{StLvl_{t-1,\tilde{t},r,c,e,m}}{1 - dis_{t,\tilde{t},r,c,e,m}}^{\text{\makebox(10,7
)[c]{\,}}}}^\text{loss adjusted previous level}  + \overbrace{in_{t,\tilde{t},r,c,e,m}+ \frac{St_{t,\tilde{t},r,c,e,m}^{in}}{1/eff_{t,\tilde{t},r,c,e,m}^{stI}}}^\text{storage inputs}  - \overbrace{\frac{St_{t,\tilde{t},r,c,e,m}^{out}}{eff_{t,\tilde{t},r,c,e,m}^{stO}}}^\text{storage outputs} \\&\hspace{-140pt}
\forall e \in \Gamma^{st}, \, c \in \gamma_{e}^{st},\,  \langle t, r \rangle \in \varphi_{c}, \, \tilde{t} \in \theta_{e,t}^{dis}, \, \xi \in \omega_{t,\tilde{t},r,e}^{st}
\label{eq:27}
\end{split}
\end{equation}
In the storage balance, $dis$ refers to the self-discharge rate, while $eff^{stI}$ and $eff^{stO}$ account for losses associated with charging and discharging. Similar to the conversion balance $\omega^{st}$ provides all sets of modes that require an individual balance. Lastly, the parameter $in$ accounts for external inputs into the storage system, for instance inflows into hydro reservoirs.

\subsubsection{Ratio constraints} \label{34}

Ratios among in- and output carriers can be restricted by an equality, greater-than or less-than constraint. Since all constraints on in- or output ratios are structured the same, only the equality constraint on output carriers is formulated in Eq. \ref{eq:28}.
\begin{align}
\underbrace{ \sum_{\langle \hat{t},\hat{r} \rangle \in \sigma_{t,r,c}} Cv_{\hat{t},\tilde{t},\hat{r},c,e,m}^{out}}_\text{output of restricted carrier} = ratio_{t,\tilde{t},r,c,e,m}^{out, eq} \underbrace{\sum_{c' \in \gamma_{e}^{out}} \; \sum_{\langle \hat{t},\hat{r} \rangle \in \sigma_{t,r,c'}}  Cv_{\hat{t},\tilde{t},\hat{r},c',e,m}^{out}}_\text{output of all carriers}  && \forall \langle t,\tilde{t},r,c,e,m \rangle \in \{ \Omega | \, ratio_{t,\tilde{t},r,c,e,m}^{out, eq} \} \label{eq:28}
\end{align}
The parameter $ratio^{out,eq}$ specifies a carrier's share of the total output. In the example, it is defined for the share of electricity in total outputs of CCGT plants with CHP, coal plants and fuel cells. Accordingly, only in these cases the corresponding constraints are created.

\subsubsection{Capacity constraints}

Dispatch variables are constrained to not exceed the operating capacities $Cap^{opr}$. To compare dispatch expressed in energy units with capacities, which are expressed in power units, dispatch variables are corrected for the length of the respective dispatch time-step. To this end we define the function $s$ that assigns a correction factor $s(t)$ for each time-step $t$. As explained in section \ref{23}, expansion can be modeled with greater spatial detail than dispatch and as a result comparing expansion with dispatch requires aggregation. For this purpose, expansion regions of technology $e$ are termed $\eta_{e}^{sp}$ and by default their resolution corresponds to the most detailed resolution across all carriers assigned, as expressed in Eq. \ref{eq:101}.
\begin{equation}
\eta_{e}^{sp} \coloneqq \{ V(R) \, | \, d(r) = \max\limits_{c \in \gamma_{e}^{in} \cup \gamma_{e}^{out}}(dep_{c}^{exp,sp}) \} \label{eq:101}
\end{equation}
Since conversion capacities transform carriers modeled at different granularities, the question arises at which resolution capacity constraints should be enforced. To answer this, part A of the appendix introduces an algorithm that determines the smallest set of constraints required for dispatch variables to comply with the operated capacities $Cap^{opr,cv}$. For each technology $e$ this set is referred to as $\Psi_{e}$ and can be split into in- and output. The corresponding constraints are provided by Eqs. \ref{eq:29} and \ref{eq:30}.
\begin{subequations} 
\begin{align}
s(t) \sum_{c \in \kappa} \; \sum_{\langle \hat{t},\hat{r} \rangle \in \sigma_{t,r,c}} \; \sum_{m \in \mu_{e}} \frac{Cv_{\hat{t},\tilde{t},\hat{r},c,e,m}^{in}}{ava_{\hat{t},\tilde{t},\hat{r},e,m}^{cv}} \leq \sum_{\hat{r} \in \eta_{e}^{sp} \cap \delta_{r}^{+}} Cap_{\alpha_{t}^{sup} ,\tilde{t},\hat{r},e}^{opr,cv} && \forall e \in \Gamma^{cv}, \, \langle \kappa,t,r\rangle  \in \Psi_{e}^{in}, \, \tilde{t} \in \theta_{e,t}^{dis} \label{eq:29}  \\
s(t) \sum_{c \in \kappa} \; \sum_{\langle \hat{t},\hat{r} \rangle \in \sigma_{t,r,c}} \sum_{m \in \mu_{e}} \frac{Cv_{\hat{t},\tilde{t},\hat{r},c,e,m}^{out}}{ava_{\hat{t},\tilde{t},\hat{r},e,m}^{cv} \, eff_{\hat{t},\tilde{t},\hat{r},e,m}^{cv}} \leq  \sum_{\hat{r} \in \eta_{e}^{sp} \cap \delta_{r}^{+}} Cap_{\alpha_{t}^{sup} ,\tilde{t},\hat{r},e}^{opr,cv} && \forall e \in \Gamma^{cv}, \, \langle \kappa,t,r\rangle \in \Psi_{e}^{out}, \, \tilde{t} \in \theta_{e,t}^{dis} \label{eq:30}
\end{align}
\end{subequations}
In the introduced formulation capacities of technologies generally refer to input capacities, which is why the constraint on output capacity in Eq. \ref{eq:30} must be corrected for the respective efficiency.

For storage, capacity constraints are separately enforced for storage input $stI$, storage output $stO$ and storage size $stS$. All storage carriers initially assigned to a technology are denoted as $\gamma_{e}^{stCap}$ and each of these carriers has individual storage capacities. Within a constraint, storage capacities for a carrier $c$ are compared with dispatch variables of all the carriers $\hat{c}$ explicitly stored. The corresponding constraints are given by Eqs. \ref{eq:31} to \ref{eq:33}.
\begin{subequations} 
\begin{align}
s(t) \sum_{\hat{c} \in \delta_{c}^{+} \cap \gamma_{e}^{st}} \; \sum_{\langle \hat{t},\hat{r} \rangle \in \sigma_{t,r,\hat{c}}} \; \sum_{m \in \mu_{e}} \frac{St_{\hat{t},\tilde{t},\hat{r} ,\hat{c},e,m}^{in}}{ava_{\hat{t},\tilde{t},\hat{r},\hat{c},e,m}^{stI}} \leq  \sum_{\hat{r} \in \eta_{e}^{sp} \cap \delta_{r}^{+}} Cap_{\alpha_{t}^{sup} ,\tilde{t},\hat{r},e,c}^{opr,stI} &&  \forall e \in \Gamma_{e}^{st},\,c \in \gamma_{e}^{stCap} , \,\langle t, r \rangle \in \varphi_{c},\, \tilde{t} \in \theta_{e,t}^{dis} \label{eq:31}\\
s(t) \sum_{\hat{c} \in \delta_{c}^{+} \cap \gamma_{e}^{st}} \; \sum_{\langle \hat{t},\hat{r} \rangle \in \sigma_{t,r,\hat{c}}} \; \sum_{m \in \mu_{e}} \frac{St_{\hat{t},\tilde{t},\hat{r} ,\hat{c},e,m}^{out}}{ava_{\hat{t},\tilde{t},\hat{r},\hat{c},e,m}^{stO}} \leq \sum_{\hat{r} \in \eta_{e}^{sp} \cap \delta_{r}^{+}} Cap_{\alpha_{t}^{sup} ,\tilde{t},\hat{r},e,c}^{opr,stO} &&  \forall e \in \Gamma_{e}^{st},\,c \in \gamma_{e}^{stCap},\, \langle t, r \rangle \in \varphi_{c} ,\, \tilde{t} \in \theta_{e,t}^{dis} \label{eq:32}\\
\sum_{\hat{c} \in \delta_{c}^{+} \cap \gamma_{e}^{st}} \; \sum_{\langle \hat{t},\hat{r} \rangle \in \sigma_{t,r,\hat{c}}} \; \sum_{m \in \mu_{e}} \frac{stLvl_{\hat{t},\tilde{t},\hat{r} ,\hat{c},e,m}}{ava_{\hat{t},\tilde{t},\hat{r},\hat{c},e,m}^{stL}} \leq \sum_{\hat{r} \in \eta_{e}^{sp} \cap \delta_{r}^{+}} Cap_{\alpha_{t}^{sup} ,\tilde{t},\hat{r},e,c}^{opr,stS} &&  \forall e \in \Gamma_{e}^{st},\,c \in \gamma_{e}^{stCap}, \, \langle t, r \rangle \in \varphi_{c},\, \tilde{t} \in \theta_{e,t}^{dis} \label{eq:33}
\end{align}
\end{subequations}
Unlike all other capacities, constraints on storage size do not include a scaling factor, because storage size already is provided in energy units.

For exchange, capacities are created for all regions and carriers defined in $\beta_{c,r}$ and capacities are then compared with dispatch variables $\hat{c}$ explicitly exchanged. Exchange capacities can be directed, meaning the energy transportable from $r$ to $r'$ and from $r'$ to $r$ can differ.
\begin{align}
s(t) \sum_{\hat{c} \in \lambda(c)} \; \sum_{\langle \hat{t},\hat{r} \rangle \in \sigma_{t,r,\hat{c}}} \; \sum_{\hat{r}' \in \rho_{\hat{c}} \cap \delta_{r'}^+ } Exc_{\hat{t},\hat{r},\hat{r}' ,c} \leq Cap_{\alpha_{t}^{sup} ,r,r',c}^{opr,exc} && \forall \langle c, \,r \rangle \in \{ V(C) \times V(R) \, | \, \beta_{c,r} \},\, r' \in \beta_{c,r}, \, t \in \tau_{c}
\end{align}

\subsubsection{Expansion}

The operated capacities $Capa^{opr}$ for conversion, storage and exchange that restrict dispatch variables do not necessarily match installed capacities $Capa^{ist}$. The framework can endogenously decide to decommission installed capacities before the end of their technical lifetime to mitigate operating costs. The following constraints achieve this for conversion capacities and are equally applicable for storage and exchange:
\begin{align}
Cap_{t,\tilde{t},r,e}^{opr,cv} \leq Cap_{t,\tilde{t},r,e}^{ist,cv} && \forall e \in \Gamma^{cv} , t \in \Phi, \tilde{t} \in \theta_{e,t}^{dis}, r \in \eta_{e}^{sp} \label{eq:36} \\
Cap_{t,\tilde{t},r,e}^{opr,cv} \leq Cap_{t-1,\tilde{t},r,e}^{opr,cv} + Exp_{t,r,e}^{cv} && \forall e \in \Gamma^{cv}, t \in \Phi, \tilde{t} \in \theta_{e,t}^{dis}, r \in \eta_{e}^{sp} \label{eq:37}
\end{align}
Eq. \ref{eq:36} simply ensures operated capacities do not exceed installed capacities. To avoid that decommissioned capacities are put into operation again, Eq. \ref{eq:37} demands that any rise in operated capacity has to result from capacity expansion, which is denoted as $Exp_{t,r,c,e}^{cv}$.

Installed capacities are a result of pre-existing capacities and capacity expansion. Analogously to expansion regions, time-steps of expansion are termed $\eta_{e}^{tp}$ and their resolution corresponds to the most detailed resolution across all carriers assigned as well:
\begin{equation}
\eta_{e}^{tp} \coloneqq \{ V(R) \, | \, d(r) = \max\limits_{c \in \gamma_{e}^{in} \cup \gamma_{e}^{out}}(dep_{c}^{exp,tp}) \}
\end{equation}
As explained in section \ref{24}, certain technologies are differentiated by time-step of construction, for others the time-step of construction is irrelevant or they cannot be expanded at all. This affects how expansion variables have to be aggregated to obtain installed capacities and is reflected by the set $\theta_{e,\tilde{t},t}^{exp}$ defined in Eq. \ref{eq:38}. The set provides all expansion time-steps to be aggregated for obtaining capacities of technology $e$ with construction period $\tilde{t}$ at time-step $t$. Consequently, this set is empty for technologies that cannot be expanded. For \textit{mature} technologies it contains all time-steps of expansion that result in a lifespan including $t$. For \textit{emerging} technologies, capacities are not aggregated and accordingly only $\tilde{t}$ itself is assigned.
\begin{equation}
\theta_{e,t,\tilde{t}}^{exp} \coloneqq \begin{cases} 
    \emptyset & \text{,if } g(e) = \textit{`stock'} \\
      \{\tilde{t}' \in \eta_{e}^{tp} \, | \, \tilde{t}' \in (\alpha_{t}^{sup} -lt_{e,\tilde{t}'} ,\alpha_{t}^{sup}] \} & \text{,if } g(e) = \textit{`mature'} \\
      \{ \tilde{t} \} & \text{,if } g(e) = \textit{`emerging'} 
   \end{cases} \label{eq:38}
\end{equation}
Building on this, in \ref{eq:39} the installed capacities are defined as the sum of expansion plus pre-existing capacities $capa^{pre}$ set exogenously. 
\begin{align}
Cap_{t,\tilde{t},r,e}^{ist,cv} = cap_{t,\tilde{t},r,e}^{pre,cv} + \sum_{\tilde{t}' \in \theta_{e,\tilde{t},t}^{exp}} Exp_{\tilde{t}',r,e}^{cv} && \forall e \in \Gamma^{cv}, t \in \Phi, \tilde{t} \in \theta_{e,t}^{dis}, r \in \eta_{e}^{sp} \label{eq:39}
\end{align}

\section{Application of the model formulation} \label{4}

To demonstrate feasibility of the presented formulation, the model its introduction was based on is now created and solved. A particular focus is on how temporal granularity impacts model size, solve time and final results. 

For this application, the open-source modelling framework AnyMOD.jl that implements the graph-based formulation is used. Code and documentation of AnyMOD.jl are freely available on Github \citep{Goeke2020b}. The corresponding repository and a Zenodo upload with all the other files to run the example model are provided in the Supplementary Material.

As introduced in the previous section, the example models the transformation of the power and heating sector from a fossil to a renewable system over the course of 20 years in 5 years steps for two stylized regions, but could be freely extended and altered. This includes the addition of energy carriers and technologies to cover more sectors or a different structure of time-steps to achieve different temporal resolutions. Also, the temporal resolution of expansion could be increased for certain technologies to model a constant expansion rate within each decade. 

\subsection{Results of the example model} 
The example model was parameterized as follows: For location-dependent parameters, like demand or availability of renewables, values were selected such that the regions \textit{East} and \textit{West} resemble Germany and France. Costs and technological properties were based on recent estimates. To actually achieve the levels of renewables and sector integration the framework was developed for, the yearly emission limit linearly decreases from 350 million tons of CO\textsubscript{2} in 2020 to zero in 2040.

\begin{figure}
	\centering
		\includegraphics[scale=.8]{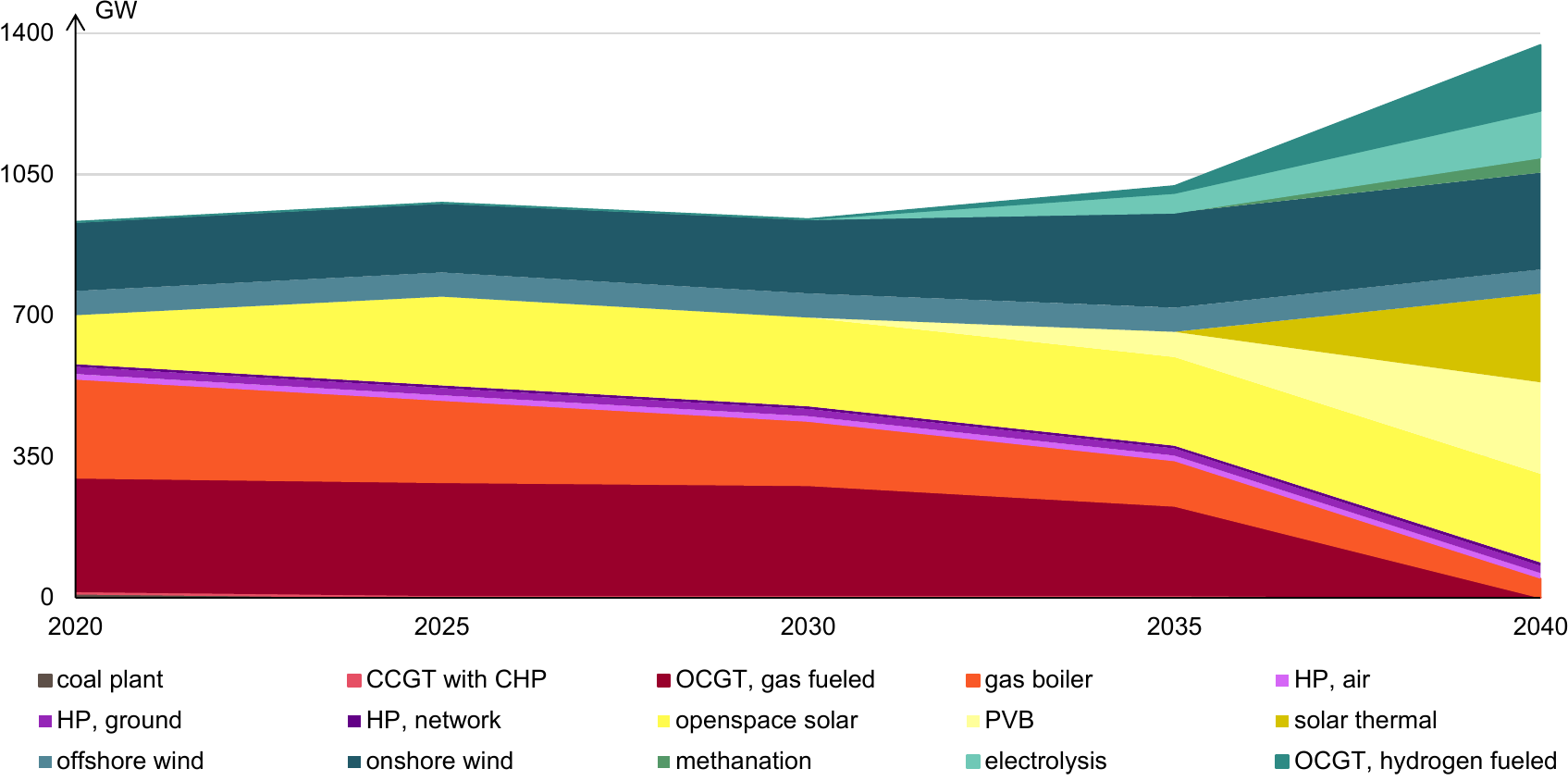}
	\caption{Operated conversion capacities for the example model}
	\label{fig:7}
\end{figure}

\begin{figure}
	\centering
		\includegraphics[scale=.75]{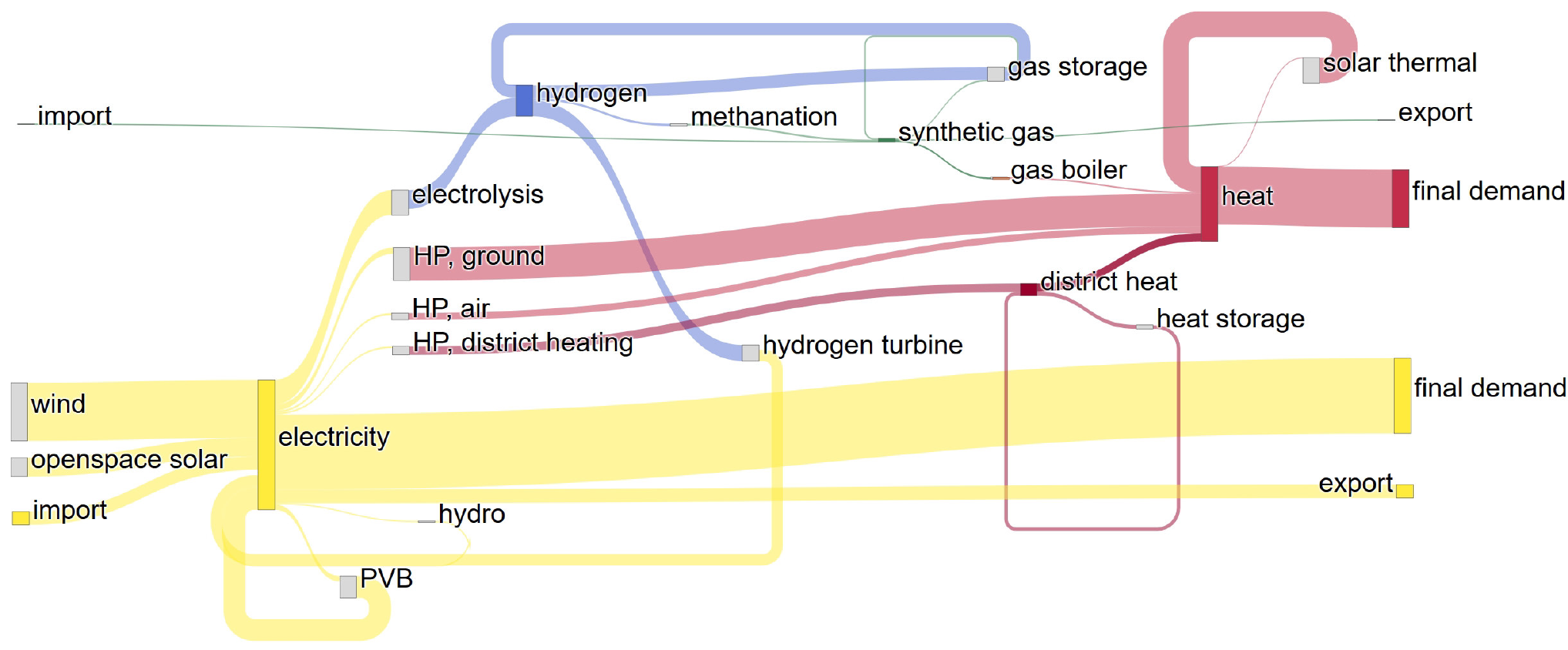}
	\caption[bla]{Quantitative energy flow in example model for all regions in the year 2040\footnotemark }
	\label{fig:8}
\end{figure}

The resulting development of operated conversion capacities is displayed in Fig. \ref{fig:7}. It should be noted that according to the framework's convention, these are input capacities. 
In the graph, the impact of moving from a small emissions limit in 2035 to no emissions in 2040 is very pronounced. Instead of switching to synthetic gas, gas boilers and OCGT power plants are mostly decommissioned and replaced with solar heating and hydrogen turbines. The resulting energy flow for 2040 is shown in Fig. \ref{fig:8}, which is the quantitative counterpart to Fig. \ref{fig:5} from section \ref{24}. Again, colored vertices represent energy carriers and grey vertices correspond to technologies. 
The graph visualizes several characteristics of the framework's graph-based approach. For example, the flow leaving \textit{district heat} and entering \textit{heat} reflects that according to the energy balance in Eq. \ref{eq:13}, descendant carriers are included in an ancestors energy balance. As a result, \textit{district heat} can equally satisfy final demand for \textit{heat} despite being produced by different technologies. Also, both \textit{hydrogen} and \textit{synthetic gas} flows enter and leave the \textit{gas storage} technology, which was defined to store their ancestor \textit{gas}. This corresponds to the storage implementation presented in sections \ref{24} and \ref{33}.

\subsection{Impact of temporal resolution} 
All these results were obtained solving the model with full foresight and the settings outlined in section \ref{23}, which proposed an hourly resolution for electricity, four-hour steps for heat and daily balancing of all gaseous carriers. To study the impact of impact temporal granularity, two more detailed scenarios are considered in addition. One extends hourly granularity to \textit{heat} and \textit{district heat}, while all other resolutions remain unchanged. In the other, all carriers are modeled with hourly resolution.

\begin{figure}
	\centering
		\includegraphics[scale=.5]{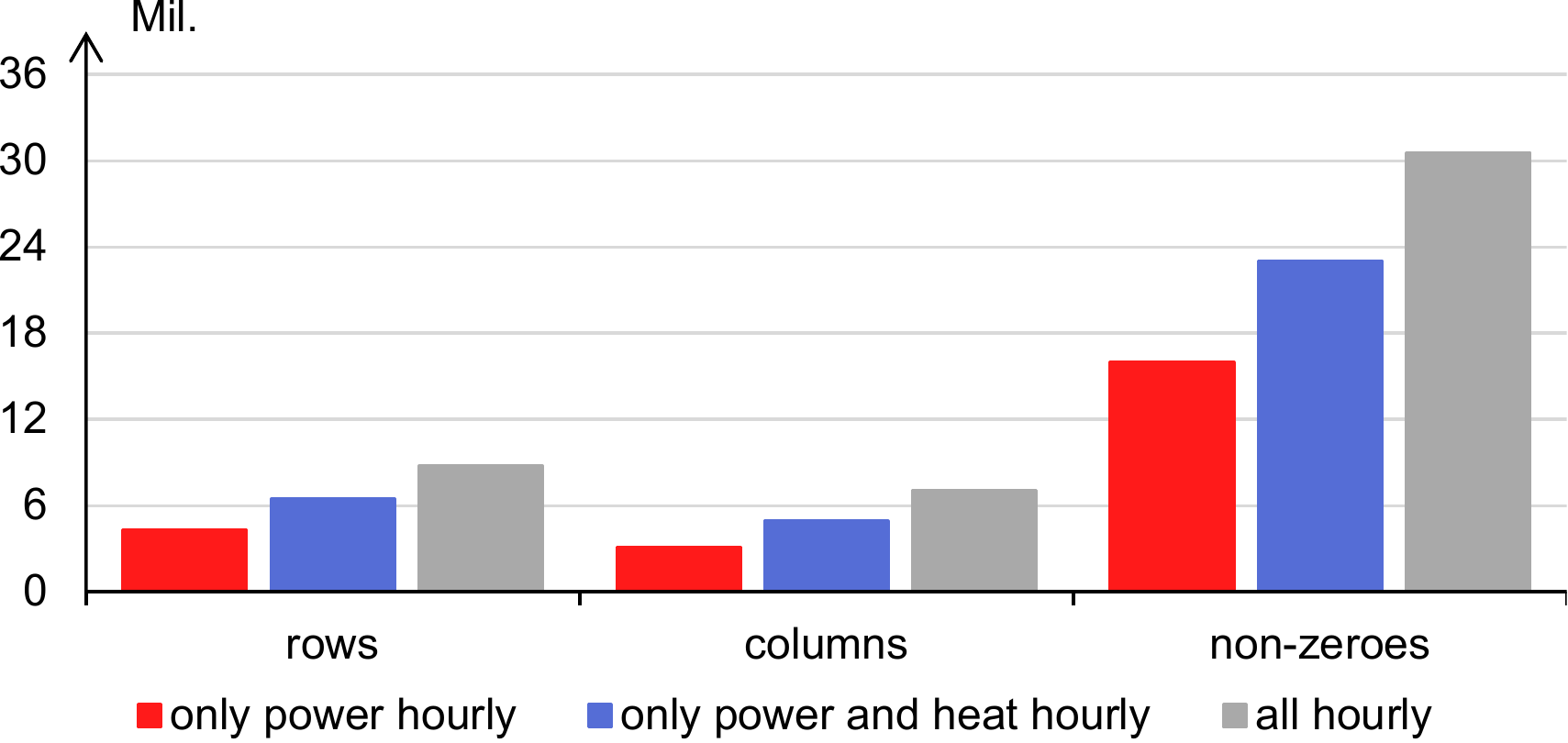}
	\caption{Model size across scenarios}
	\label{fig:9}
\end{figure}

In Fig. \ref{fig:9} the size and number of non-zero elements for the model's underlying optimization matrix are shown across all three scenarios. Even though three-quarters of technologies in the model either use or generate electricity, reducing temporal granularity for all carriers but electricity achieves a reduction of about 50\% in matrix size and number of non-zero elements. If resolution for heat and electricity is kept hourly and detail is only decreased for gaseous carriers, the reduction still amounts to 25\%.
A reduced model size will decrease working memory requirements and makes it possible to solve models that previously did not fit into memory, but it does not necessarily reduce computation time. The time to solve a problem also depends on the inner structure of the matrix and the applied solution algorithm. 

\footnotetext{Import and export flows are aggregated across all regions, and thus have the same value.}

\begin{figure}
	\centering
		\includegraphics[scale=.6]{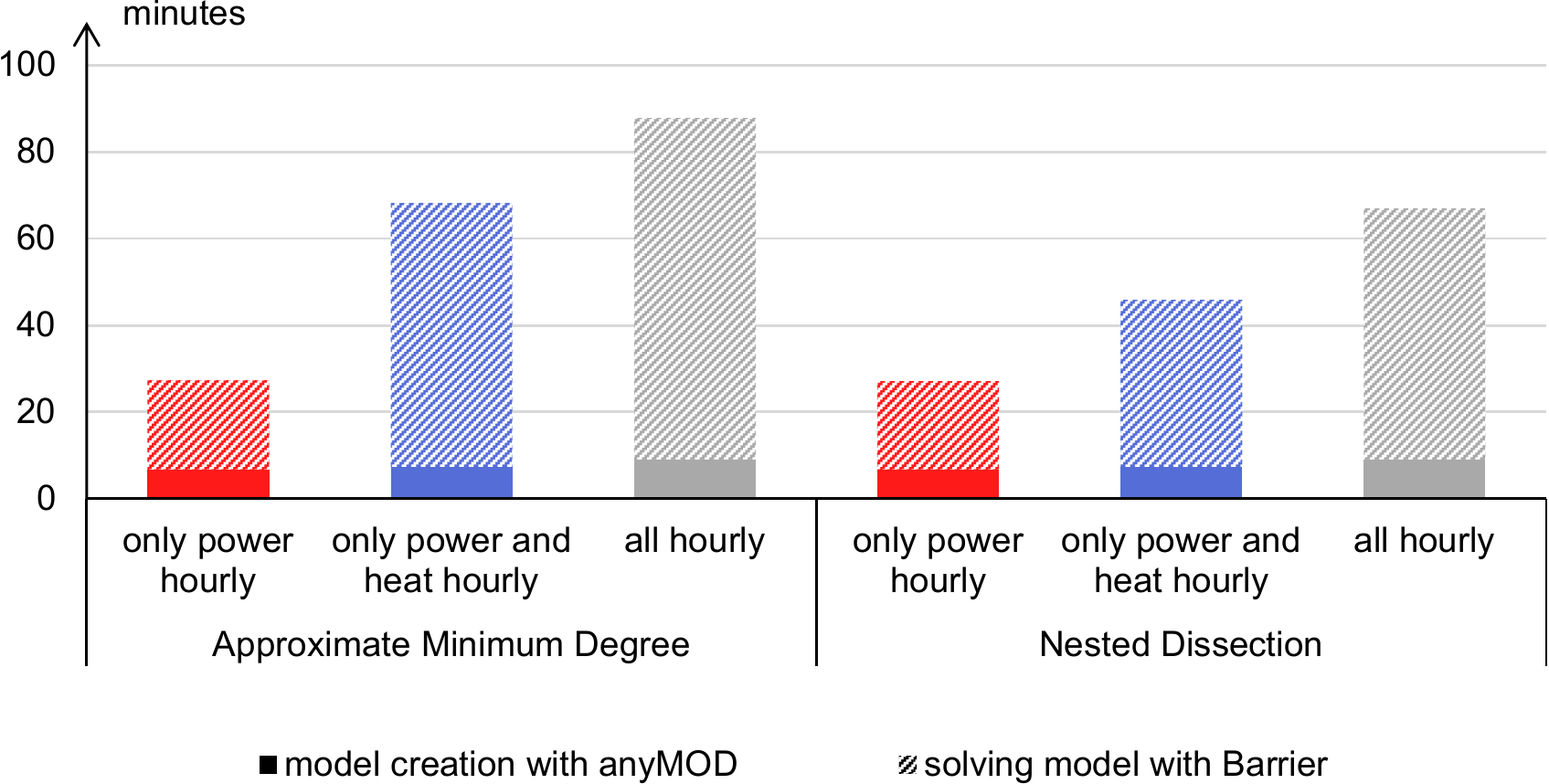}
	\caption{Solve time with Barrier algorithm across scenarios}
	\label{fig:10}
\end{figure}

To assess the scenarios in terms of computation time, they were solved using different algorithms of the Gurobi solver. Using the simplex method did not provide any results in less than a day; solve times when applying the Barrier algorithm with 'Approximate Minimum Degree' or 'Nested Dissection' ordering are displayed in Fig. \ref{fig:10}.\footnote{Reported times only refer to the barrier algorithm itself and omit crossover. In no case crossover improved results by more than \num{0.000007} percent, but typically increased computation time by a factor of four.}
Results indicate that solve time decreases disproportionately to model size. When going from an hourly granularity for all carriers to only modeling electricity hourly, model size was reduced by 50\%, but solve time decreased by 64\% to 75\% depending on the ordering method. The corresponding computations were run on a high-performance computing cluster. If reproduced on a desktop computer with less working memory and parallel processors, the model creation might take longer, because the framework heavily utilizes multi-threading. Also, for 'Nested Dissection' ordering, memory limits are likely to be exceeded.

\begin{figure}
	\centering
		\includegraphics[scale=.8]{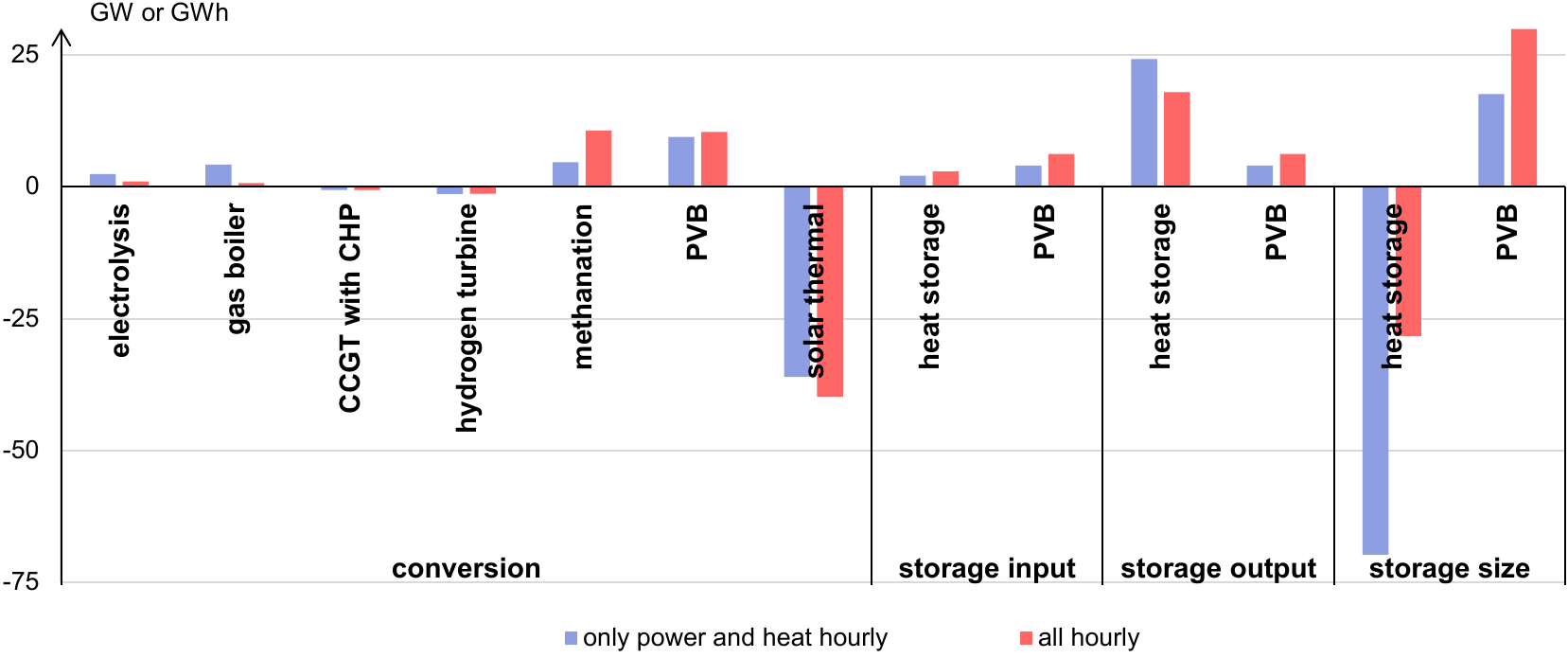}
	\caption{Operated capacities compared to reference case in 2040}
	\label{fig:11}
\end{figure}

Lastly, final model results are compared for the three scenarios. To this end, Fig. \ref{fig:11} shows the difference in operated capacities for the two more detailed scenarios compared to the reference case for 2040. Positive values indicate that capacities for the more detailed scenario exceed results from the reference case. Only technologies where results differ are included.
If heat is modeled with hourly resolution, generation from CHP plants and solar heating is partly replaced by more flexible gas boilers fueled by synthetic gas. To generate this gas, additional capacities for electrolysis and methanation are required. CHP plants generating less leads to smaller sized heat storage. Also, reduced solar thermal capacity allow the installation of additional PVB systems, since both technologies compete for rooftop area. If the resolution of gas is changed from daily to hourly as well, shifting gas within the day requires gas storage and thus becomes subject to losses. Consequently, storing gas is avoided and instead methanation capacities are increased to produce gas when required. For the reference case system costs amount to 397.4 € billion and increase to 399.6 € billion when heat is additionally modeled at an hourly resolution. Modeling all carriers hourly further increases costs to 400.1 € billion.

Deviations between the reference case and more detailed scenarios should not necessarily be interpreted as inaccuracies. If a less detailed resolution can be justified from an engineering perspective, it does not only reduce model size, but also allows the consideration of the system's inherent flexibility. Consequently, the decrease in system costs when reducing a carrier's granularity can be interpreted as the economic value of this flexibility. The effects that changing the granularity of a single carrier has across the entire system also emphasizes what was stated at the very beginning of the introduction: Analyzing energy systems characterized by high shares of intermittent renewables requires a cross-sectoral perspective.

\section{Conclusion and outlook} \label{5}

This paper introduced a novel formulation for modelling macro-energy systems. In contrast to existing formulation, it pursues a novel approach based on graph theory. Organizing sets in rooted trees enables two features that facilitate modeling systems with high shares of renewables and sector integration. First, the method allows the level of temporal and spatial detail to be varied by energy carrier. As a result, model size can be reduced without reducing the level of detail applied to fluctuating renewables. In addition, flexibility inherent to the system, for example in the gas network, can be accounted for. Second, substitution of energy carriers can be modeled in dependence of the respective context: conversion, storage, transport, or demand. This achieves a more comprehensive representation of how technologies and energy carriers can interact in an integrated energy system. In addition, smaller features not found in previous frameworks, namely an accurate representation of technological advancement, endogenous decommissioning and internal storage of generated carriers, have been implemented.

To demonstrate its capabilities, the graph-based formulation was applied to a stylized example that models the transformation of the power and heating sector from a fossil to a renewable system over the course of 20 years in two regions loosely based on Germany and France. The example shows in particular how varying the temporal resolution by carrier reduces solve time by 64\% to 75\% without imposing a major bias on results.  

So far, the introduced formulation cannot account for weather related uncertainties of renewable generation, although this has been identified as a key requirement for modeling high shares of renewables \citep{Ringkjob2018}. Therefore, the focus of further development is to enable stochastic capacity expansion to account for a range of weather years. Since this implies a substantial increase in model size, a particular challenge lies in solving such models. One approach could be to implement a distributed solution algorithm based on Benders decomposition that can fully exploit the capabilities of high-performance computing \citep{Conejo2006}.

\section*{Acknowledgements}

The research leading to these results has received funding from the European Union's Horizon 2020 research and innovation program under grant agreement No 773406. Also, I want to thank Thomas Heggarty and Mario Kendziorski for their constructive feedback on earlier drafts of this paper. A special thanks goes to Mario Kendziorski and Richard Weinhold for their help with the Julia programming language. 

\section*{Supplementary material}

Code and documentation of the AnyMOD.jl framework can found in the following GitHub repository: \href{https://github.com/leonardgoeke/AnyMOD.jl}{https://github.com/leonardgoeke/AnyMOD.jl}. All other scripts and data files to run the example are available on Zenodo (\url{https://doi.org/10.5281/zenodo.4699276}). The upload also includes additional information on the input parameters used. The example uses version 0.1.0 of AnyMOD.jl.

\appendix

\numberwithin{equation}{section}
\makeatletter 
\newcommand{\section@cntformat}{Appendix \thesection:\ }
\makeatother

\section{Set of required capacity constraints}
The algorithm to obtain the smallest set of constraints required to correctly restrict the use of conversion capacities is displayed below. The key part is of the algorithm is carried out separately for input and output carriers and for the temporal and spatial domain. In the first step, the respective input and output carriers are obtained and sorted according to their temporal or spatial depth. In case the respective technology requires a conversion balance, carriers that are inputs and have the same granularity as the conversion balance can be omitted from further analysis. In these cases, the conversion balance itself already ensures correct use of the installed capacities. The algorithm then iterates over the remaining carriers. Within this iteration, $\kappa$ is used to collect the current carrier $c$ and all carriers of previous iterations. For each iteration, the spatial or temporal depth of the current carrier $c$, the smallest depth among all carriers in $\kappa$, and $\kappa$ itself are written to the set $\psi_{e}$. When this has been done for the temporal and spatial dimension, redundant entries are removed from $\psi_{e}$. An entry is redundant, if it includes the same or less carriers then another entry, but is not more detailed, neither in the temporal nor spatial domain.

For the output of CCGT plants with CHP for example, the resulting temporal and spatial sets of $\psi$ are provided by Eqs. \ref{eq:201} and \ref{eq:202}.
    \begin{align}
\begin{split} \label{eq:201} 
    \psi_{e}^{out,tp} = & \{ \{\{'electricity'\},5,1\}, \\
    & \{\{'electricity','district heat'\},4,1\}\} \} 
\end{split}\\
\begin{split} \label{eq:202}
    \psi_{e}^{out,sp} = & \{ \{\{'district heat'\},4,2\}, \\
    & \{\{'district heat','electricity'\},4,1\}\} \}
\end{split}
\end{align}
The second set of $\psi_{e}^{out,sp}$ or $\psi_{e}^{out,tp}$ is redundant and can be removed. The remaining entries of $\psi_{e}^{out}$ are then used to create the set $\Psi_{e}^{out}$ replacing depths with actual time-steps and regions. The only input carrier of CCGT plants is \textit{gas} which is modeled at the same resolution as the conversion balance. Therefore, it is removed within the algorithm, $\Psi_{e}^{in}$ is empty and no capacity constraint on input variables must be enforced in this case. \\

\setcounter{equation}{0}
\SetArgSty{textnormal}
\begin{algorithm*}[H]
\SetAlgoLined
 \For{\textit{'input'/'output'}}{
   \For{\textit{'temporal'/'spatial'}}{
        \uIf{\textit{'input'}}{
            $\upsilon = \gamma_{e}^{use}$\;}
        \ElseIf{\textit{'output'}}{
            $\upsilon = \gamma_{e}^{gen}$\;}
       sort $\upsilon$ ascending by $dep_{c}^{dis,tp/sp}$\;
        \If {\textit{'input'} $\textbf{and}$ $\gamma_{e}^{use} \neq \emptyset$ \textbf{and} $\gamma_{e}^{gen} \neq \emptyset$}{
        filter $c$ with $dep_{c}^{dis,tp/sp} = \min\limits_{c \in \gamma_{e}^{gen} \cup \gamma_{e}^{use}}dep_{c}^{dis,tp/sp}$ from $\upsilon$\;}
        $\kappa = \emptyset$\;
      \For{$c \in \upsilon$}{
          $\kappa = \kappa \cup \{c\}$\;
          \uIf{\textit{'temporal'}}{
            add $\langle \kappa,\, dep_{c}^{dis,tp},\, \min\limits_{c' \in \kappa}{(dep_{c}^{dis,sp})} \rangle$ to $\psi_{e}$\;
          }
          \ElseIf{\textit{'spatial'}}{
            add $\langle \kappa,\, \min\limits_{c' \in \kappa}{(dep_{c}^{dis,tp})},\, dep_{c}^{dis,tp} \rangle$ to $\psi_{e}$\;
          }}}
    filter redundant entries of $\psi_{e}$\;
    $\Psi_{e}^{in/out} = \{ \langle \kappa, z^{tp},z^{sp} \rangle  \in \psi \, | \, \{\kappa\} \times \{ V(T) \, | \, d(t) = z^{tp}\} \times \{ V(R) \, | \, d(r) = z^{sp}\}\}$;}
\caption{Determine constraints on conversion capacity for technology $e$}
\end{algorithm*}

\section{Objective function and limiting constraints}
\setcounter{equation}{0}

The frameworks objective function given in Eq. \ref{eq:1010} minimizes costs. These are compromised of expansion costs $Cost^{exp}$, operating costs $Cost^{opr}$, variable costs $Cost^{var}$ and trade costs $Cost^{trd}$.
\begin{equation}
    \min Cost^{trd} + \sum_{t \in \Phi} Cost_{t}^{exp} + Cost_{t}^{opr} + Cost_{t}^{var} \label{eq:1010}
\end{equation}
Expansion costs includes costs for expanding conversion, exchange, storage-input and storage-output capacities as well as costs related to storage size:
\begin{align}
    Cost_{t}^{exp} = Cost_{t}^{exp, cv} + Cost_{t}^{exp, stI} + Cost_{t}^{exp, stO} + Cost_{t}^{exp, stS} + Cost_{t}^{exp, exc} && \forall t \in \Phi
\end{align}
Each of these cost components is computed by summing the product of the discount factor $disc$, the annuity $ann$, and the expansion variable $Exp$ for each time-step in a technologies lifetime. Operating costs are obtained analogously, but instead of the annuity and expansion variable, operating costs $opr$ are multiplied with the installed capacities $Capa^{opr}$. In Eqs. \ref{eq:102} and \ref{eq:103} both equations are exemplary provided for conversion capacities.
\begin{align}
Cost_{t}^{exp, cv} = \sum_{e \in \Gamma^{cv}} \; \sum_{r \in \eta_{e}^{sp}} \; \sum_{\tilde{t} \in [t,t+ lt_{e,\tilde{t}})} disc_{t,r}\; ann_{\tilde{t} ,r,e}^{cv}\;Exp_{\tilde{t},r,e}^{cv} && \forall t \in \Phi \label{eq:102} \\
Cost_{t}^{opr , cv} = \sum_{e \in \Gamma^{cv}} \; \sum_{r \in \eta_{e}^{sp}} \; \sum_{\tilde{t} \in [t,t+ lt_{e,\tilde{t}}) } disc_{t,r}\; opr_{\tilde{t} ,r,e}^{cv}\;Capa_{t,\tilde{t},r,e}^{opr,cv} && \forall t \in \Phi \label{eq:103}
\end{align}
Variable cost can be imposed on all used, generated, charged, discharged or exchanged quantities:
\begin{align}    
    Cost_{t}^{var} = Cost_{t}^{var, use} + Cost_{t}^{var, gen} + Cost_{t}^{var, stI} + Cost_{t}^{var, stO} + Cost_{t}^{var, exc} && \forall t \in \Phi
\end{align}
The corresponding constraints are only created where a corresponding cost parameter $var$ is defined. In Eq. \ref{eq:104} this is expressed for used quantities. In the example model, this is only the case for quantities used by the methanation technology to account for the carbon the process requires. 
\begin{align}
    Cost_{t}^{var, use} = \sum_{\langle t,\tilde{t},r,c,e,m \rangle \in \{ \Omega \, | \, var_{t,\tilde{t},r,c,e,m}^{use} \}} disc_{t,r} \; var_{t,\tilde{t},r,c,e,m}^{use}\; Use_{t,\tilde{t},r,c,e,m} && \forall t \in \Phi \label{eq:104}
\end{align}
Trade costs reflect the costs or revenues from trading energy with exogenous markets.  Prices on these markets are reflected by the parameter $prc$ and the entire costs can be defined by equation \ref{eq:105}.
\begin{align}
    Cost^{trd} = \sum_{c \in V(C)} \; \sum_{r \in \rho_{c}} \; \sum_{t \in \tau_{c}} disc_{t,r} \; ( \sum_{i \in \zeta^{buy}} prc_{t,r,c,i}^{buy} \;Trd_{t,r,c,i}^{buy} - \sum_{i \in \zeta^{sell}} prc_{t,r,c,i}^{sell} \;Trd_{t,r,c,i}^{sell}) \label{eq:105}
\end{align}

Similar to constraints on input or output ratios in section \ref{34}, the creation of limiting constraints depends on the dimension of the provided parameters. If limits are defined at a resolution less detailed then the corresponding variables, constraints apply to the sum of all descendant variables. For example, the limit on installed capacities in Eq. \ref{eq:106} can be defined for the vertex \textit{rooftop} in the rooted trees of technology. As a result, an upper limit will be enforced on the installed capacities of the descendant technologies \textit{PVB} and \textit{solar thermal}.
\begin{align}
    capaL_{t,r,e}^{inst, cv} = \sum_{\hat{t} \in \delta_{t}^{+}} \; \sum_{\hat{r} \in \delta_{r}^{+}} \; \sum_{\hat{e} \in \delta_{e}^{+}} \; \sum_{\tilde{t}' \in \theta_{e,t,\tilde{t}}^{exp}} Capa_{\hat{t},\tilde{t}',\hat{r},\hat{e}}^{inst,cv} && \forall \langle t,r,e \rangle \in \{ \langle t,r,e \rangle | \,  capaL_{t,r,e}^{inst, cv}\} \label{eq:106}
\end{align}
A special case are emission constraints, because they are not applied to variables, but to the product of the emission factor $emF$ and used quantities $Use$ as denoted in Eq. \ref{eq:107}.\footnote{Optionally, the emission constraint can be extended to additionally account for storage and exchange losses.} 
\begin{align}
    emL_{t,r} = \sum_{\hat{t} \in \delta_{t}^{+}} \; \sum_{\hat{r} \in \delta_{r}^{+}} \;  \sum_{\langle \hat{t},\tilde{t},\hat{r},c,e,m \rangle \in \{ \Omega \, | \, emF_{\hat{t},\tilde{t},\hat{r},c,e,m}^{use} \}} emF_{\hat{t},\tilde{t},\hat{r},c,e,m} \; Use_{\hat{t},\tilde{t},\hat{r},c,e,m} 
    && \forall  \langle t,\tilde{t},r,c,e,m \rangle \in \{ \Omega | \, emL_{t,\tilde{t},r,c,e,m}^{out, eq} \} \label{eq:107}
\end{align}

\printcredits

\bibliographystyle{model3-num-names.bst}

\bibliography{cas-refs}
\end{document}